\documentclass[journal=aapmcd,manuscript=article,layout=twocolumn]{achemso}
\usepackage[english]{babel}
\usepackage[utf8]{inputenc}
\usepackage[colorinlistoftodos, color=green!40, prependcaption]{todonotes}
\usepackage{esvect}
\usepackage{xcolor}
\usepackage{soul}

\usepackage{epstopdf}
\AppendGraphicsExtensions{.tif}

\usepackage{achemso}
\setkeys{acs}{maxauthors = 0}

\usepackage{float}

\usepackage{upgreek}
\usepackage{subcaption}
\usepackage[pdftex, pdftitle={Article}, pdfauthor={Author}]{hyperref} 
\usepackage{totalcount}
\usepackage{indentfirst}

\author{Damian Bouwmeester}%
\email{d.bouwmeester@tudelft.nl}
\affiliation{%
Kavli Institute of Nanoscience, Delft University of Technology, Lorentzweg 1, 2628 CJ Delft, The Netherlands
}%

\author{Talieh S. Ghiasi}
\affiliation{%
Kavli Institute of Nanoscience, Delft University of Technology, Lorentzweg 1, 2628 CJ Delft, The Netherlands
}%

\author{Gabriela Borin Barin}
\affiliation{%
nanotech@surfaces Laboratory, Empa, Swiss Federal Laboratories for Materials Science and Technology, 8600 Dübendorf, Switzerland
}%

\author{Klaus M{\"u}llen}
\affiliation{Max Planck Institute for Polymer Research, 55128 Mainz, Germany
}%

\author{Pascal Ruffieux}
\affiliation{%
nanotech@surfaces Laboratory, Empa, Swiss Federal Laboratories for Materials Science and Technology, 8600 Dübendorf, Switzerland
}%

\author{Roman Fasel}
\affiliation{%
nanotech@surfaces Laboratory, Empa, Swiss Federal Laboratories for Materials Science and Technology, 8600 Dübendorf, Switzerland
}%
\affiliation{Department of Chemistry, Biochemistry and Pharmaceutical Chemistry, University of Bern, Freiestrasse 3, CH-3012 Bern, Switzerland}

\author{Herre S.J. van der Zant}
\affiliation{%
Kavli Institute of Nanoscience, Delft University of Technology, Lorentzweg 1, 2628 CJ Delft, The Netherlands
}%

\date{\today}


\title{
MoRe Electrodes with 10-nm Nanogaps for Electrical Contact to Atomically Precise Graphene Nanoribbons}

\begin{document}




\begin{abstract}

Atomically precise graphene nanoribbons (GNRs) are predicted to exhibit exceptional edge-related properties, such as localized edge states, spin polarization, and half-metallicity.
However, the absence of low-resistance nano-scale electrical contacts to the GNRs hinders harnessing their properties in field-effect transistors. In this paper, we make electrical contact with 9-atom-wide armchair GNRs using superconducting alloy MoRe as well as Pd (as a reference), which are two of the metals providing low-resistance contacts to carbon nanotubes. We take a step towards contacting a single GNR by fabrication of electrodes with a needle-like geometry, with about 20~nm tip diameter and $10$~nm separation. To preserve the nano-scale geometry of the contacts, we develop a PMMA-assisted technique to transfer the GNRs onto the pre-patterned electrodes. Our device characterizations as a function of bias-voltage and temperature, show a thermally-activated gate-tunable conductance in the GNR-MoRe-based transistors. 

\end{abstract}

\maketitle

\begin{figure*}
     \centering

    \includegraphics[width=\linewidth]{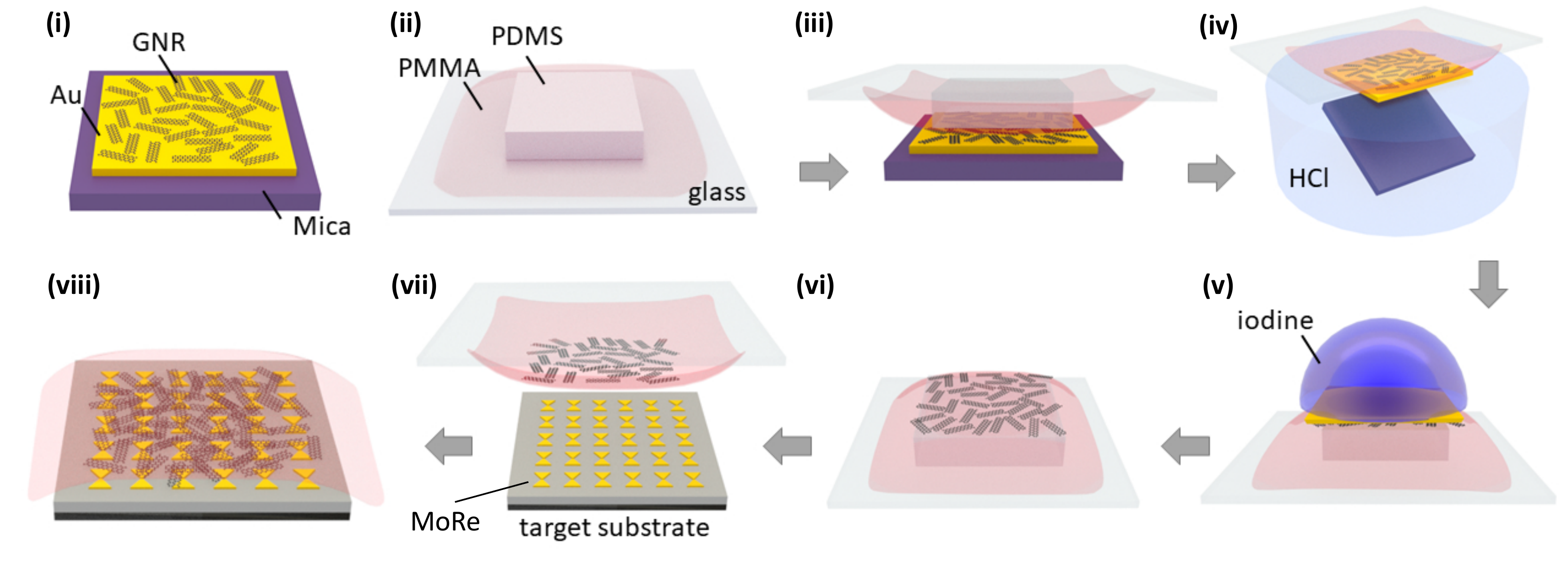}

    \caption{PMMA membrane GNR transfer method. (i) Au(111) on mica 9-AGNR growth substrate. (ii) PMMA membrane on PDMS placed on a glass slide. (iii) Aligned placement of the PMMA membrane onto the GNR growth substrate. The stage is heated to 130~$^\circ$C to promote stronger adhesion of the PMMA membrane to the GNR growth substrate. (iv) Delamination of the mica layer in 30\% HCl. (v) Gold etching in KI/I$_2$ for 5 to 10 minutes followed by rinsing and soaking overnight in DI water. (vi) GNRs on PMMA membrane after drying in ambient conditions (vii) Stamping of the PMMA membrane onto the target substrate, followed by annealing at 150~$^\circ$C. (viii) Target substrate with PMMA-covered GNRs after GNR transfer}
    \label{fig:1}
\end{figure*}


\section*{Introduction}
Graphene nanoribbons (GNRs) are a quasi-1D analogue of graphene. Although graphene is classified as a 2D semimetal, a combination of quantum confinement and electron-electron interactions make GNRs semiconducting~\cite{Nakada1996,Son2006_2,Yang2007}. The electronic band gap of GNRs scales inversely with their width and depends on their edge structure~\cite{Son2006_2}. This tunable band gap is a property that is of interest for field-effect transistors (FETs)~\cite{Schwierz2010,Llinas2017} and optoelectronics~\cite{Miao2021}. Moreover, GNRs exhibit edges and ends that can host localized spin-polarized states~\cite{Nakada1996,Son2006,Tao2011} which is interesting for spintronic applications. Since the electronic properties of GNRs are sensitive to their width and edge structure, edge disorder and width variation at the atomic level result in hopping transport within the ribbon~\cite{Martin2009}, which significantly suppresses their intrinsic electronic/spintronic properties~\cite{Brede2023}. 

On-surface, bottom-up chemical approaches have enabled the synthesis of a variety of GNRs from molecular precursors with structural precision on the atomic level~\cite{Cai2010},  such as armchair~\cite{Cai2010}, zigzag~\cite{Ruffieux2016}, chevron~\cite{Cai2010}, staggered/topological~\cite{Groning2018,Rizzo2018}, metallic~\cite{Rizzo2020} and substitutionally doped GNRs~\cite{Bronner2013}. The electronic properties of these GNRs have mostly been studied by scanning tunneling microscopy. Characterization of the intrinsic electronic properties of the GNRs in electronic circuits has been comparatively limited by high contact resistances, limited chemical stability of the edges, and short GNR length below typical source-drain contact distances. 

Low-resistance Ohmic electrical contacts are important for obtaining large on-state currents in FET devices, as well as for studying the intrinsic transport properties of GNRs. However, making consistent, low-resistance electrical contact to atomically-precise GNRs is challenging due to their on-surface synthesis with typical 1-nm width that leads to large band gaps (of the order or $1$~eV). Their typical length scale also results in a small contact area, of the order of $10$~$\mathrm{nm}^2$. Recently, there have been considerable efforts to electrically contact atomically precise armchair-edge GNRs by electrodes made of graphene~\cite{Martini2019,Elabbassi2020,Braun2021,Zhang2022_2}, carbon nanotubes(CNTs)~\cite{Zhang2022}, palladium~\cite{Llinas2017,Mutlu2021, BorinBarin2022,Lin2023} and gold~\cite{Richter2020}. These studies, however, are still limited compared to the detailed characterization of a large variety of contacts to CNTs~\cite{Chen2005,Svensson2011}. Even though CNTs structurally differ from GNRs by the absence of edges, small diameter (less than $1.0$~nm) CNTs are the closest system to atomically precise GNRs due to their similar band structure and considerable band gap (larger than $0.8$~eV)~\cite{Matsuda2010}. 

For the case of semiconducting CNTs, Schottky barriers are formed at the metal-CNT interfaces, the size of which depends on the chosen contact metal and the diameter of the nanotube. The presence of Schottky barriers results in a contact resistance that increases as the temperature is decreased. For CNTs, a distinction is often made between physisorption and chemisorption~\cite{Zhang2000,Sarkar2014}, as well as p-type (high work function) and n-type (low work function) electrical contacts~\cite{Svensson2011}. Typically, n-type contacts form with metals that are prone to oxidation (Al, Sc, Y, Ti), while p-type contacts can be made with noble metals (Au, Pt, Pd), as well as Ni, Co, Mo and W.

Two of the metals that stand out for making low-resistance electrical contacts with small or absent Schottky barriers to CNTs are Pd~\cite{Javey2003,Mann2003} and Mo~\cite{Cao2015}/MoRe alloy~\cite{Schneider2012,Kaikkonen2020}. Pd contacts to 9- and 13-atom-wide armchair GNRs~(9-AGNRs and 13-AGNRs) have already been studied in a short-channel FET geometry by Llinas et al.~\cite{Llinas2017}, who found that transport in their devices was limited by tunneling through a Schottky barrier at the contacts. Nevertheless, their Pd-contacted 9-AGNR FETs with a $\textrm{high-}\upkappa$~HfO$_2$ gate dielectric have a large on-state current ($>1$~$\upmu$A), as well as an on-off ratio of $10^5$. Mo/MoRe on the other hand is of interest as it is a superconducting metal, which may be used to induce superconductivity in GNRs by the superconducting proximity effect~\cite{Meissner1960} at cryogenic temperatures. \textcolor{black}{In a weakly transparent electrical contact, the superconducting energy gap can be used to perform tunneling spectroscopy of the GNRs, while a highly transparent contact would allow for utilizing GNRs in Josephson junctions.}

Here, we further explore MoRe and Pd contacts to 9-AGNRs by studying their current-voltage characteristics at various temperatures. In particular, we compare two distinct electrode geometries that have the potential to respectively address many GNRs in parallel, and single GNRs.  
With the aim of contacting single 9-AGNRs, \textcolor{black}{an electrode design was made that minimizes gap width.} Here we fabricate needle-like MoRe and Pd nanogap electrodes with a minimum width of $\sim20$~nm and spacing of 6-15~nm. \textcolor{black}{The smaller gap spacing achieved for this geometry could also enable addressing shorter GNRs.} \textcolor{black}{The polymer-free transfer method was attempted on this geometry, resulting in broken MoRe nanowires due to etching}. In order to preserve the \textcolor{black}{ more fragile needle-like} nanogaps and the contact geometry from etchants used in polymer-free GNR-transfer recipes, here we develop a PMMA-membrane-assisted technique for the transfer of the 9-AGNR films, based on the PMMA fishing transfer technique introduced by Martini et al.~\cite{Martini2019}. This technique keeps the electrodes intact by preventing direct contact with any liquid and allows for controlled handling and $\sim1~\upmu$m precise placement of the GNRs onto the electrical contacts using micro-manipulators. Our transfer method offers the additional advantage of \textcolor{black}{using} \textcolor{black}{a stretched and clamped} PMMA film, which \textcolor{black}{could reduce} wrinkling and folding. \textcolor{black}{With this technique, we fabricate 10-nm nanogap MoRe and Pd devices and we investigate and compare their performance. We show that the 10-nm Pd nanogap devices have a few orders of magnitude higher conductance, which suggests that a Pd/MoRe bilayer thin film would be a better contact material for the realization of functional superconducting GNR devices.}


\section*{Results}

The 9-AGNRs were grown by on-surface synthesis\textcolor{black}{\cite{DiGiovannantonio2018}}, discussed in detail in the methods section. The average length of the 9-AGNRs used in this work is 45~nm.

The two distinct electrode geometries used here to address the GNRs are the wide-nanogap and needle-like geometries. The wide-nanogap geometry consists of a pair of $2~\upmu$m long parallel wires, separated by approximately 30~nm. This geometry was made to address transport through many GNRs in parallel. The needle-like nanogap geometry consists of two opposing nanowires that are cuspated at a 30$^\circ$ angle, separated by less than 15~nm. This geometry minimizes the contact area and thus, increases the chance of making contact with a single GNR. The fabrication of these two electrode geometries is discussed in detail in the methods section.

Prior to the GNR transfer, the nanogap electrodes were characterized by recording the current versus bias voltage ($IV$ characteristic) in the bias range of -1~to~1~V. Only devices that were found to be electrically open (resistance $\geq$ 1~T$\Omega$ at 1~V) were used in this study. The transfer of the GNRs onto wide MoRe nanogaps was performed by a polymer-free method~\cite{BorinBarin2019}. For the needle-like MoRe and Pd nanogap devices, we resorted to a polymethylmethacrylate (PMMA) membrane-based transfer method, because the gold etchant destroys the MoRe and Pd nanowires. The procedure for making the PMMA membrane for the GNR transfer is detailed in the methods section.


As this PMMA-membrane transfer method has not yet been applied to GNRs, we discuss it in detail, following the steps illustrated in Fig.~\ref{fig:1}. In the first step, the PMMA-PDMS stamp held on a glass slide (ii) is brought in contact with the GNR film, grown on a Au-mica substrate (i), using micromanipulators of a transfer stage, shown in (iii). After the contact of the PMMA membrane and the GNR film, the stage is heated to 130~$^\circ$C to promote stronger adhesion. The stack of PDMS-PMMA-GNR-Au(111)-mica held on the glass slide is then put into 30\% HCl until the mica is delaminated from the Au film, 
depicted in (iv). The glass slide is then rinsed and soaked in DI water three times, before leaving it to dry in ambient conditions. As shown in (v), KI/I$_2$ gold etchant is next drop-casted onto the Au film with a pipette and left for 10 minutes to fully etch the 200~nm Au film. The GNR-PMMA-PDMS stack is then rinsed and soaked in DI water overnight to remove residual iodine stains. After drying in ambient conditions(vi), the PMMA membrane is perforated around the PDMS using a needle to allow for its easier detachment from the PDMS stamp in the next step. The PMMA-GNR film is brought into contact with the pre-fabricated electrodes at the transfer stage, heated up to 150~$^\circ$C to improve adhesion (vii). In the final step, the glass slide-PDMS stamp is retracted, leaving the PMMA-covered GNR film on the electrodes (viii). After the transfer, the devices were post-annealed for 30 minutes at 150~$^\circ$C to reflow the PMMA layer which would increase the chance of making better contact with the GNR film.

\begin{figure*}
    \centering
    \begin{subfigure}[]{0.3\linewidth}
        \phantomcaption{}
        \centering
        \includegraphics[scale=1]{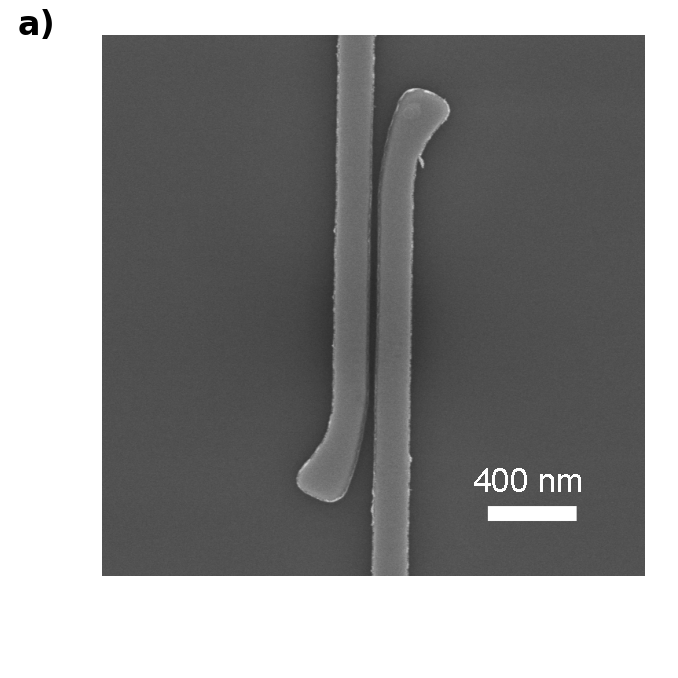}

        \label{fig:2a}
    \end{subfigure}
    \begin{subfigure}[]{0.3\linewidth}
        \phantomcaption{}
        \centering
        \includegraphics[scale=1]{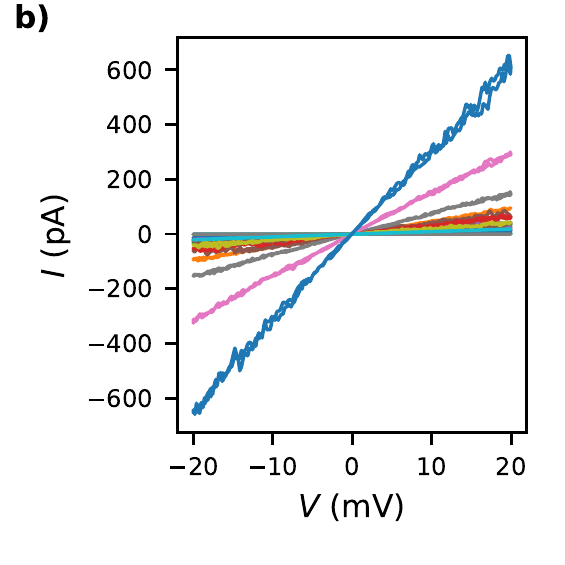}
        
        \label{fig:2b}
    \end{subfigure}
    \begin{subfigure}[]{0.3\linewidth}
        \phantomcaption{}
        \centering
        \includegraphics[scale=1]{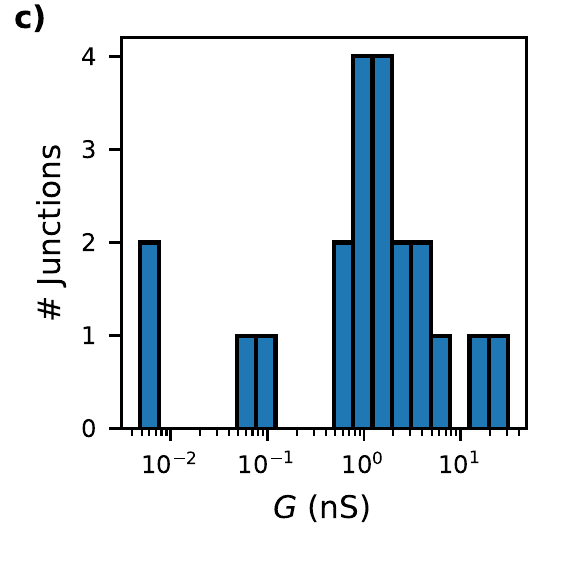}

        \label{fig:2c}
    \end{subfigure}
    \begin{subfigure}[b]{0.3\linewidth}
        \phantomcaption{}
        \centering
        \includegraphics[scale=1]{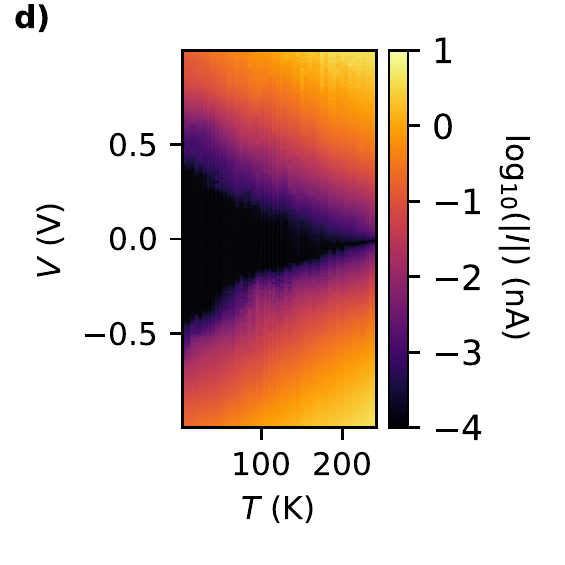}
        
        \label{fig:2d}
    \end{subfigure}
    \begin{subfigure}[b]{0.3\linewidth}
        \phantomcaption{}
        \centering
        \includegraphics[scale=1]{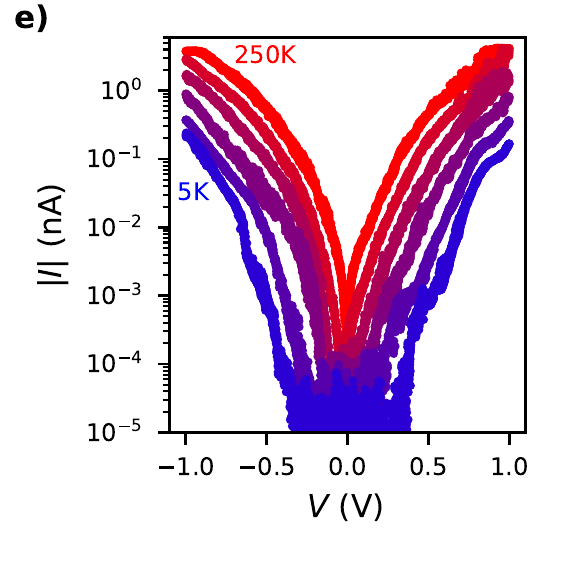}
        
        \label{fig:2e}
    \end{subfigure}
    \begin{subfigure}[b]{0.3\linewidth}
        \phantomcaption{}
        \centering
        \includegraphics[scale=1]{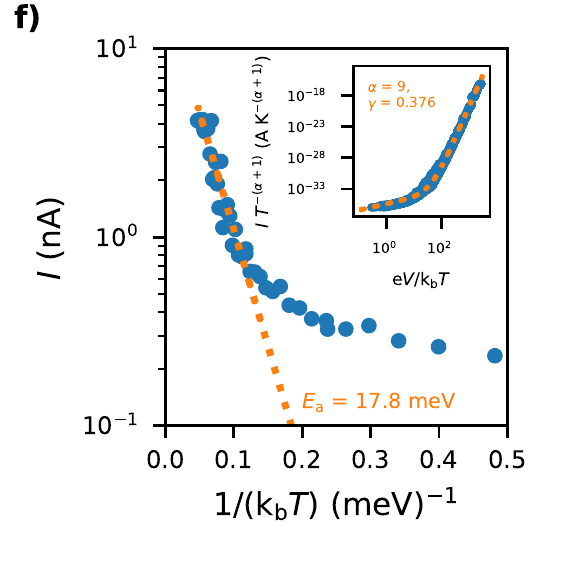}
        
        \label{fig:2f}
    \end{subfigure}

    \caption{Room temperature characterization. a) SEM image of a 2~$\upmu$m wide MoRe nanogap contact electrodes; The electrode seperation is 30~nm and the scalebar is 400~nm.
    b) $IV$ curves of wide MoRe nanogap 9-AGNR devices. \textcolor{black}{Each color corresponds to a measurement performed on a different device.}
    c) Histogram of conductances of wide MoRe nanogap 9-AGNR devices.
    d) Map of current versus bias voltage and temperature of a selected wide MoRe nanogap 9-AGNR device.
    e) Corresponding temperature dependence of the $IV$ characteristic extracted from d) for $T$ = 5, 50, 100, 150, 200, and 250~K.
    f) Corresponding temperature dependence of the current at $V$ = 1~V extracted from d). The inset shows the rescaled curve with a guide to the eye based on the nuclear tunneling model.
    }
    \label{fig:2}
\end{figure*}

\begin{figure*}
    \centering
    
     \begin{subfigure}[]{0.3\linewidth}
        \phantomcaption{}
        \centering
        \includegraphics[scale=1]{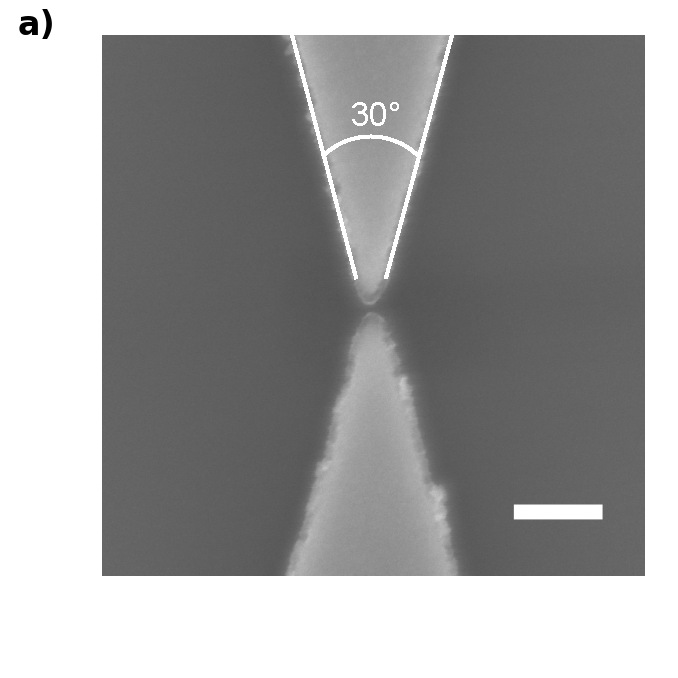}

        \label{fig:3a}
    \end{subfigure}
    \begin{subfigure}[]{0.3\linewidth}
        \phantomcaption{}
        \centering
        \includegraphics[scale=1]{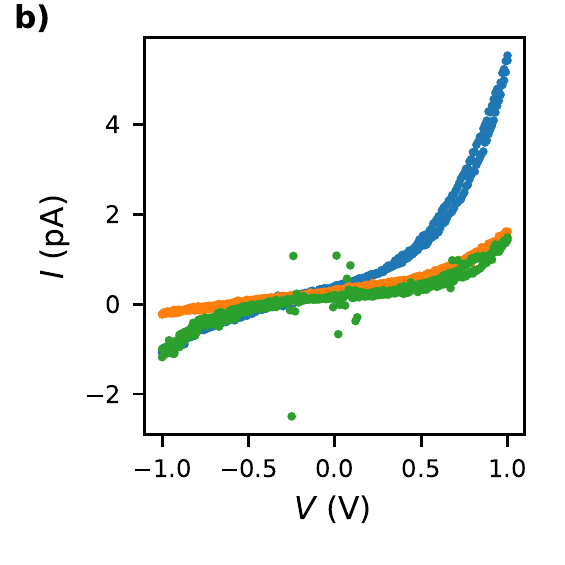}
         
        \label{fig:3b}
    \end{subfigure}
    \begin{subfigure}[]{0.3\linewidth}
        \phantomcaption{}
        \centering
        \includegraphics[scale=1]{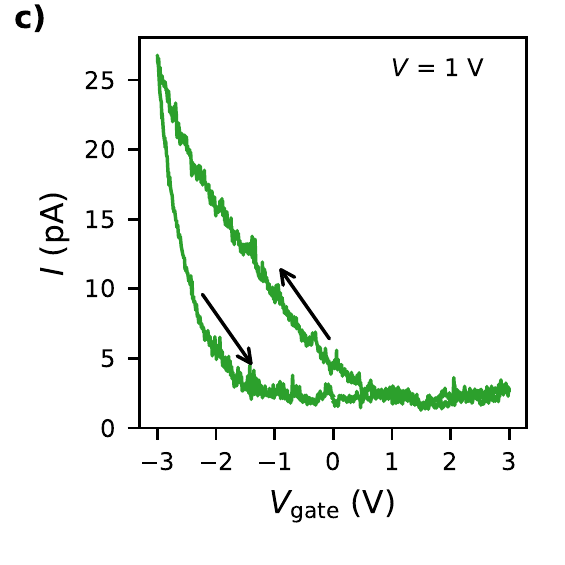}
         
        \label{fig:3c}
    \end{subfigure}
    \begin{subfigure}[b]{0.3\linewidth}
        \phantomcaption{}
        \centering
        \includegraphics[scale=1]{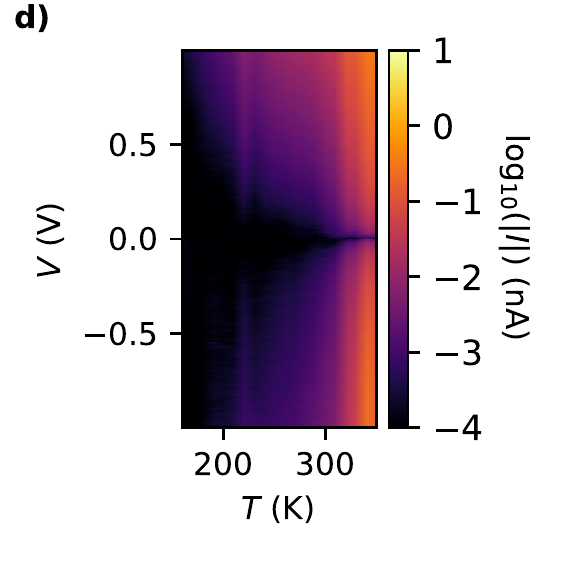}
        
        \label{fig:3d}
    \end{subfigure}
    \begin{subfigure}[b]{0.3\linewidth}
        \phantomcaption{}
        \centering
        \includegraphics[scale=1]{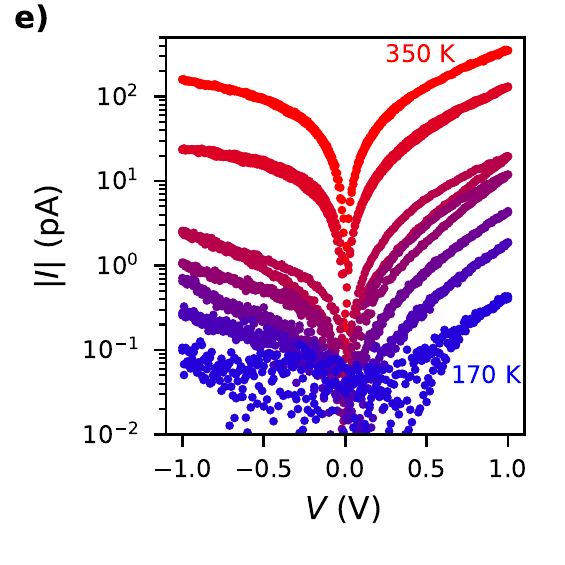}
        
        \label{fig:3e}
    \end{subfigure}
    \begin{subfigure}[b]{0.3\linewidth}
        \phantomcaption{}
        \centering
        \includegraphics[scale=1]{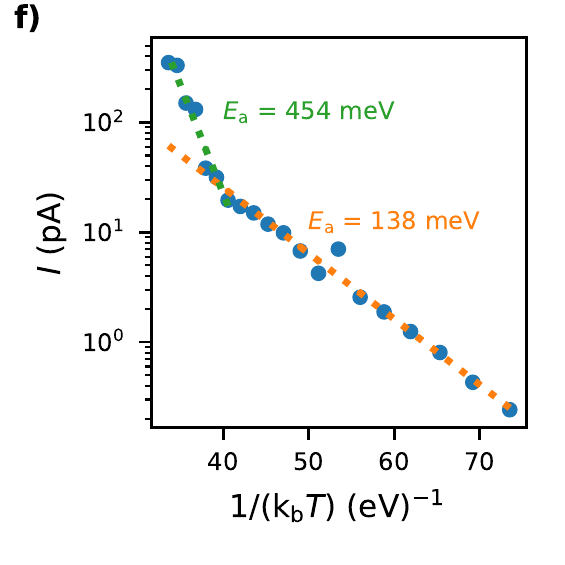}

        \label{fig:3f}
    \end{subfigure}
     
    \caption{
    a) SEM image of a representative needle-like MoRe 10-nm nanogap electrodes with a separation of roughly 10~nm; The scalebar is 100~nm.
    b) $IV$ curves of three MoRe 10-nm nanogap 9-AGNR devices. \textcolor{black}{Each color corresponds to a measurement performed on a different device.}
    c) Gate voltage dependence of the current at $V$ = 1~V of a selected MoRe 10-nm nanogap 9-AGNR device. The arrows indicate the sweep direction of the gate voltage.
    d) Map of the current versus bias voltage and temperature of the same device as in c) at a fixed gate voltage of -2~V.
    e) Corresponding temperature dependence of the $IV$ characteristic extracted from d), shown for $T$ = 170, 200, 230, 260, 290, 320 and 350~K.
    f) Corresponding temperature dependence of the current at $V$ = 1~V extracted from d). The inset shows the rescaled curve with a guide to the eye based on the nuclear tunneling model.
    }
    \label{fig:3}
\end{figure*}

\begin{figure*}
    \centering
    \includegraphics[scale=1]{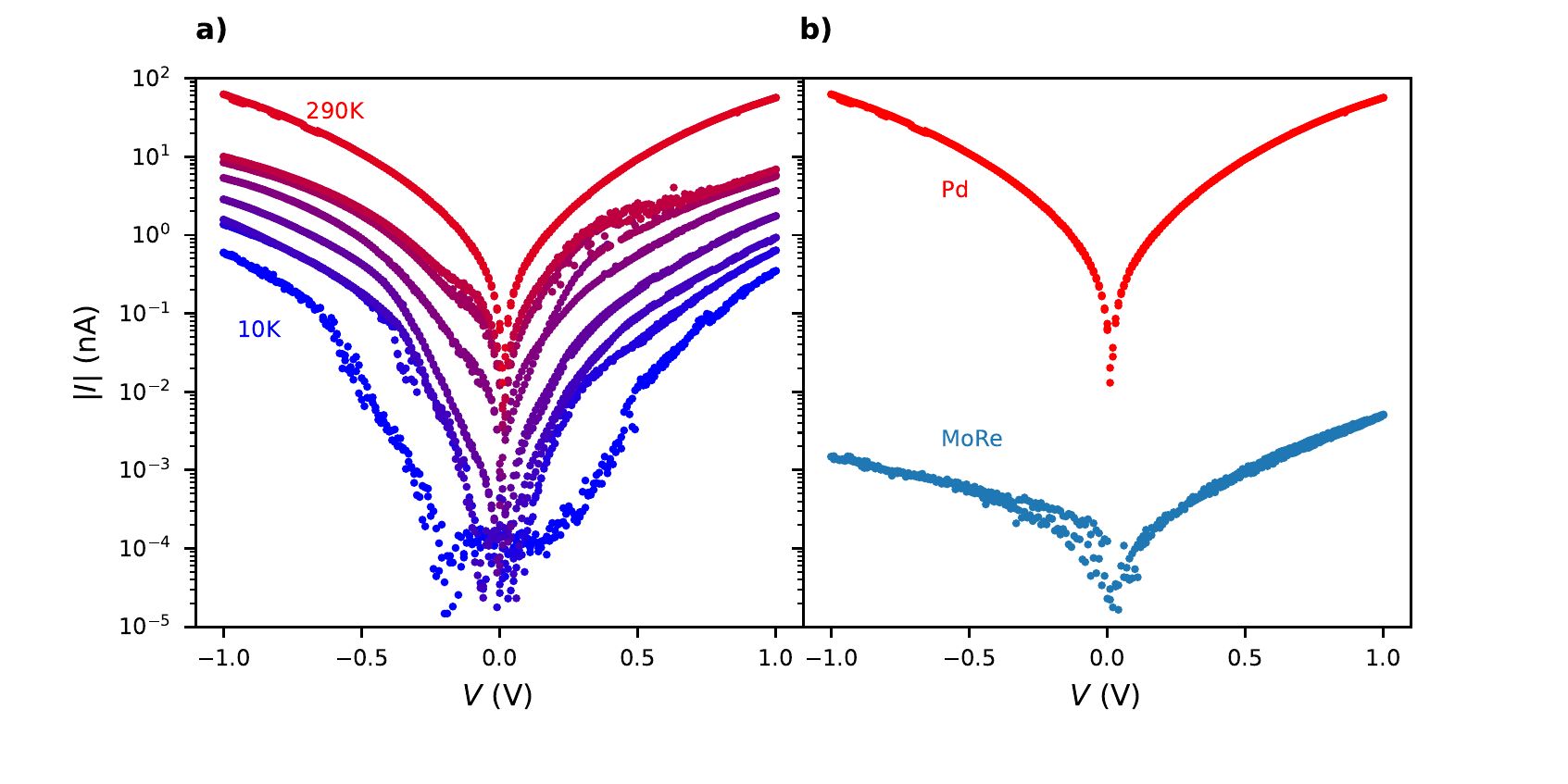}
    \caption{a) Temperature dependence of the $IV$ characteristic of a 10-nm Pd nanogap 9-AGNR device.
    b) Comparison of the $IV$ curve of the most conductive 10-nm MoRe nanogap, shown as the blue $IV$ curve in figure~\ref{fig:3b}, and the $IV$ curve of the 10-nm Pd nanogap in Fig.~4a at room temperature.}
    \label{fig:4_v2}
\end{figure*}

We first discuss the measurements of wide MoRe nanogap 9-AGNR devices, at room temperature in vacuum. \textcolor{black}{In Fig.~\ref{fig:2a} we show a representative SEM image of the wide MoRe nanogap electrodes. For all devices, }$IV$ curves were recorded in the bias range from -20 to 20~mV, which we show \textcolor{black}{together} in Fig.~\ref{fig:2b}. Out of the 22 devices onto which the GNR transfer was performed, 21 were found to be conductive. All $IV$ curves are linear within the applied bias range, with varying slopes. The electrical conductance ($G$) of the devices was extracted by fitting the slope with a linear fit, resulting in the histogram of Fig.~\ref{fig:2c}. The majority of the devices show conductance in the range of 0.5-8~nS, with a median of 1.3~nS. The standard deviation of $\mathrm{log}(G)$ is 0.95, equivalent to a standard deviation in the conductance of $\sim$1 order of magnitude. There are, however, also devices with a conductance smaller than 10~pS or larger than 10~nS, spanning over four orders of magnitude in total.

The temperature dependence of the $IV$ characteristics of one of the wide MoRe nanogaps-GNR junctions is shown in Fig.~\ref{fig:2d} on a logarithmic scale (see SI, section \textcolor{black}{1, for measurements of another device}. Furthermore an $IV$ curve up to 4V taken at 100 mK can be found in the SI, section \textcolor{black}{2}). 
The measured current at a fixed bias voltage decreases with decreasing temperature, with a kink observed at $80$~K. In Fig.~\ref{fig:2e}, the $IV$ curves plotted for various temperatures highlight the plateau-like feature at $T<100$~K, together with an increase in the slope of the logarithm of the $IV$ curve with decreasing temperature. In Fig.~\ref{fig:2f}, the current at a bias voltage of 1~V is plotted versus inverse temperature. In the high-temperature regime, an exponential decay is found, which can be described by:
\begin{equation}
I\ =\ C\ e^{-\frac{E_{\mathrm{a}}}{k_{\mathrm{B}}T}},
\label{eq:Boltzmann}
\end{equation}
where $C$ is a constant, $k_{\mathrm{B}}$ is the Boltzmann constant and $E_{\mathrm{a}}$ is an activation energy. Below $\frac{1}{k_{\mathrm{B}}T}=0.1$~$(\mathrm{meV})^{-1}$, equivalent to $T~>~$80~K, the temperature-dependence fit yields $E_{\mathrm{a}}~=~17.8$~meV.

At lower temperatures, the log(current)-voltage curve does not follow the exponential decay and instead flattens off. For this part, the scaling analysis done by Richter et al.~\cite{Richter2020} for charge transport in 9-AGNR networks was followed, which is based on a nuclear tunneling~\cite{Fisher1985,Grabert1985,Bockrath1999} model. \textcolor{black}{We note that an analysis based on the Richardson-Schottky and Simmons model was also attempted. The bias voltage dependence could be fit, but we could not simultaneously reproduce the temperature scaling. The nuclear tunneling model instead gives an efficient description of the current versus bias voltage and temperature.} The equation describing the $IV$ characteristic in this model is: 
\begin{equation}
I\ =\ I_0\ T^{\alpha+1}\ \mathrm{sinh}(\frac{\gamma e V}{2 k_{\mathrm{B}} T})\ |\Gamma (1+\frac{\alpha}{2} + i \frac{\gamma e V}{2\pi k_{\mathrm{B}} T})|^2,
\label{eq:Luttinger}
\end{equation}
where $I_0$ is a constant, $\alpha$ is a dimensionless dissipation coefficient, $\gamma < 1$ is the inverse of the number of hopping sites/voltage divisions and $\Gamma$ is the gamma function. 
In the inset of Fig.~\ref{fig:2f}, a rescaling of the data in Fig.~\ref{fig:2d} has been made by plotting $\frac{I}{T^{\alpha+1}}$ versus $\frac{eV}{k_{\mathrm{B}}T}$ on a log-log scale.
$\alpha=9$ was taken as a fixed parameter in the model to compare with the results from Richter et al~\cite{Richter2020}. When $\alpha$ is fixed, $\gamma$ determines the transition from a thermally-activated regime, where $\frac{I}{T^{\alpha+1}}$ is proportional to $\frac{eV}{k_{\mathrm{B}}T}$, to a bias-driven regime, where $\frac{I}{T^{\upalpha+1}}$ scales as $(\frac{eV}{k_{\mathrm{B}}T})^{\alpha+1}$. A guide to the eye is plotted for $\gamma = 0.378$, which shows reasonable agreement with the data. This corresponds to a voltage division over an average of roughly $3$ segments within a distance of 30~nm.

9-AGNR devices with needle-like 10-nm MoRe nanogaps are also characterized at room temperature in vacuum. \textcolor{black}{In Fig.~\ref{fig:3a} we show a representative SEM image of the needle-like MoRe electrodes.} The GNR film was transferred onto 32 pre-fabricated nanogaps. The current is measured versus the applied bias voltage up to 1~V\textcolor{black}{. }\textcolor{black}{$IV$ curves of the conductive devices are} shown \textcolor{black}{together} in Fig.~\ref{fig:3b}. Three devices are found to have a current above the noise level within this range. The $IV$ curves are nonlinear with a maximum current of 1 to 5~pA. The electrical characteristics of the two devices shown by the blue and orange curves display an asymmetry in current versus the bias-voltage which is \textcolor{black}{mostly} independent of the polarity of the source-drain contacts \textcolor{black}{and can be explained by a capacitive coupling of the source and drain electrodes to the GNRs} (see SI, section \textcolor{black}{3}). The current through the junctions was also characterized versus local bottom-gate voltage ($V_\mathrm{gate}$) at a fixed bias voltage of $V$~=~1~V. This is shown in Fig.~\ref{fig:3c} for the device represented by the green $IV$ curve in Fig.~\ref{fig:3b}. The sweep directions of the gate voltage are indicated by the arrows. The observed hysteresis (additional results on time dependence in SI, section \textcolor{black}{4}) in the trace and retrace gate-sweeps is similar to what has been reported for 7-AGNR devices at room temperature in air~\cite{Bennett2013}, as well as in vacuum for 5-AGNR and 9-AGNR devices for temperatures between 5 and 262~K~\cite{Tries2020}. Furthermore, the current is the largest at negative gate voltages, which is indicative of p-type behavior of the 9-AGNR FETs. 

Similar to the measurements on the wide-nanogap geometry, the $IV$ characteristic of the 10-nm nanogaps is measured versus temperature (measurements of another device in SI, section \textcolor{black}{1}). At a fixed $V_\mathrm{gate}~=~-2~V$, $IV$ curves were first measured cooling down from 290~K to 160~K, below which the current gets smaller than the noise floor of 100~fA. To obtain additional information, $IV$ curves are also measured warming up from 290~K to 350~K. The resulting curves are shown in Fig.~\ref{fig:3d} as a colormap, with a few individual traces in Fig.~\ref{fig:3e}. As observed in the wide-nanogap MoRe GNR devices, the current decreases by orders of magnitude as the temperature is lowered. In the needle-like MoRe devices, the asymmetry of the current with bias voltage also increases as the temperature is decreased. In contrast to what was observed in the wide-nanogap geometry in Fig.~\ref{fig:2e}, the slope of the log(current)-voltage characteristic is smaller in the 10-nm nanogap devices. The temperature dependence of the current at a bias voltage of 1~V versus inverse temperature is plotted in Fig.~\ref{fig:3f}. Activation energies are once again extracted by fitting the current versus temperature at $V =$~1~V using equation \ref{eq:Boltzmann}, which results in $E_{\mathrm{a}}=138$~meV for $T < 290$~K and $E_{\mathrm{a}}=454$~meV for $T > 290$~K. The obtained $E_{\mathrm{a}}$ for the MoRe 9-AGNR devices are an order of magnitude larger in 10-nm nanogap than those found for the wide-nanogap geometry.

Finally, we measured the $IV$ characteristic of 9-AGNR devices with 10-nm nanogap geometry made of the larger work function noble metal Pd as a function of temperature. Pd is well-known for making low-resistance Ohmic electrical contacts to CNTs, as well as GNRs~\cite{Llinas2017}, which makes it an excellent metal to compare with MoRe as a reference. Out of the 32 devices onto which 9-AGNRs were transferred, 2 were found to be electrically conductive. The $IV$ curves at various temperatures for one device are shown in Fig.~\ref{fig:4_v2}a (measurements of another device in SI, section \textcolor{black}{1}. Furthermore a nuclear tunneling scaling analysis can be found in the SI, section \textcolor{black}{5}). The $IV$ characteristic of the selected Pd device is bias symmetric. The slope of the log(I)-V characteristic is small, but larger than that measured for the MoRe 10-nm nanogaps. The devices with Pd contacts show a weaker temperature dependence of the $IV$ characteristics than those with MoRe contacts (shown in Fig.~\ref{fig:3e}). At $T=10$~K, the current is still orders of magnitude above the noise level, with currents up to $300$~pA at $1$~V bias voltage. In Fig.~\ref{fig:4_v2}b the $IV$ characteristic of the 10-nm Pd nanogap 9-AGNR device is compared with that of the most conductive MoRe nanogap of the same geometry, at room temperature. The conductance of the MoRe devices is four orders of magnitude smaller than the Pd devices at a bias voltage of $1$~V. 

\section*{Discussion}

In the $IV$ characteristics of wide-nanogap MoRe 9-AGNR devices, one thing that stands out is the variation in conductance by 4 orders of magnitude. \textcolor{black}{Some variation is expected due to inherent uncertainty in the transfer method. The GNRs are not globally aligned and may have lost local alignment during the transfer. The numbers of GNRs present in the different devices varies due to the inherent randomness of GNR film growth and positioning on the devices. Improvements can be made with regard to the effect of GNR alignment by growing and transferring globally aligned GNRs on Au (788) surfaces onto devices. In this case the alignment can be also be monitored by polarized Raman spectroscopy\cite{Overbeck2019}. For 5-AGNR devices, aligned growth on Au (788) has been shown to significantly improve the device yield and conductance\cite{BorinBarin2022}.} \textcolor{black}{The variation in device conductance over orders of magnitude however points towards} large variation in \textcolor{black}{GNR} conductance. This could possibly be explained by oxidation or inhomogeneity of the sputtered MoRe alloy contact, which would results in additional tunnel barriers and spatially varying work functions. Another possible explanation is that the gold etchant introduces a spatially non-uniform doping profile over the devices, resulting in variable band alignment. 

Another peculiarity is the kink at 80~K in Fig.~\ref{fig:2d}, which is suggestive of a change in the transport regime. The kink occurs at a voltage of approximately $\pm$200~mV, for which $eV/k_\mathrm{b}T\approx200~\mathrm{meV}~/~6.6~\mathrm{meV}\approx30$. This corresponds to the kink in the nuclear tunneling scaling plot in the inset of Fig.~\ref{fig:2f}. In the context of the nuclear tunneling model, this suggests that the kink may be a transition from a thermally dominated regime to a bias voltage-driven regime. 

The fact that the temperature and voltage dependence of the wide-nanogap MoRe 9-AGNR devices can be described using a nuclear tunneling model is surprising, considering the fact that the contact spacing is smaller than the average length of the GNRs. The earlier study by Richter et al.~\cite{Richter2020} described a possible hopping process from ribbon to ribbon. This suggests that either the dominant transport path is through roughly $3$ GNRs or that the hopping process occurs within single GNRs with a length scale between $30~\mathrm{nm}~/~3 = 10$~nm and $45~\mathrm{nm}~/~3 = 15$~nm. The hopping sites may be local trap states, in which case the sub-sections in the GNR itself act as the charge transport barrier. Such trap states could be present due to a multitude of possible causes, such as overlapping of the GNRs due to rearrangement during substrate transfer, iodine doping after the conventional wet transfer technique used for the wide-nanogaps, as well as adsorbates or charge puddles~\cite{Zhang2009,Mayamei2020}.
If the transport through our wide-nanogap MoRe devices is dominated by trap sites, that may also explain the large variability in resistances. A sparse trap density with random placement/barrier widths could result in a large variation in conductance.

\textcolor{black}{For the needle-like devices, the expected device yield, based on random angular alignment only, is $30 ^{\circ} / 180 ^{\circ}=1/6 \approx 16\%$. This is significantly larger than the observed yield, which suggests that there is a mechanism that decreases the observed device yield. We identify two possible mechanisms that could reduce device yield. Firstly, the probability of a GNR to bridge the electrodes could be reduced by rearrangements of the GNR film. Clustering/stacking of GNRs, variations in GNR density or the dissolution of GNRs into the PMMA layer and contact defects such as surface oxidation could reduce the device yield (See SI, section 6, for the Raman spectra on MoRe). Secondly, the observed device yield can be lower than the number of devices containing GNRs. We suspect the latter could be the case as the measured devices are close to the lower limit of measurable conductances in our setup ($\sim0.1$~pS)}

The $IV$ characteristic of the needle-like MoRe nanogap 9-AGNR devices display a significantly smaller slope of the log(current)-voltage characteristic. The most likely explanation for this is the gate voltage of $-2$~V, applied to improve the signal-to-noise ratio. \textcolor{black}{In SI section 7}, we show that the normalized $IV$ curves get increasingly more linear as the gate voltage goes from $0$ to $-4$~V. Another possible reason for the difference in linearity is that the dominant transport mechanism for these junctions is different, as transport can occur over a source-drain distance of only $6$~nm. This distance is smaller than the \textcolor{black}{segment length} found for the wide-nanogap MoRe devices.

From the temperature dependence of the $IV$ characteristic at $1$~V, the extracted $E_{\mathrm{a}}$ can be related to the band alignment of the contact metal with the valence band of the GNR at p-type contacts. A lower $E_{\mathrm{a}}$ implies better band alignment and a smaller Schottky barrier. The $E_{\mathrm{a}}$ extracted for the MoRe-based 9-AGNR devices are an order of magnitude larger in 10-nm nanogaps compared with that in wide nanogaps. 
Note that for these devices, the used geometry, gate dielectric, and transfer methods are different.

We believe it is unlikely that the geometry itself plays a significant role in the band alignment on the metal surface. It is possible that the high-$\upkappa$ HfO$_{2}$ and Pt local gate influence the band alignment if the GNR-metal contact is close to the oxide interface. The effective vacuum ($\upepsilon_r=1$) distance between the Pt and the GNR is about $0.5$ - $1$~nm. For CNTs embedded in Pt, an expected p-doping of $0.75$~eV was found by ab-initio calculations using a density functional theory + non-equilibrium Green’s function simulation~\cite{Fediai2016}. Assuming a similar doping effect for GNRs over a larger distance than the van der Waals gap of $0.33$~nm~\cite{Sutter2009}, the expected doping is of the order of $0.33~\mathrm{nm}~/~1~\mathrm{nm} \times (0.75~\mathrm{eV})=240$~meV, which is in magnitude comparable to the observed mismatch in activation energies. This would however result in a better valence band alignment of the GNRs in the 10-nm nanogap devices and lower activation energies, which is the opposite of what was observed. Thus this explanation based on geometry can not explain our observations. 

The transfer method may also influence the MoRe/GNR contact by means of doping. In particular, the gold etchant step, which produces iodine complexes, is known to result in p-type doping of graphene~\cite{Yao2019}. \textcolor{black}{Nanoparticle gold residues can also potentially introduce n-doping\cite{Wu2012}.} In the PMMA-membrane-based transfer technique, no contact of the gold etchant with the electrode was made and a considerably longer rinse and soak time, 8 to 24 hours versus 5 minutes, was used after the gold etchant step. This could have resulted in a lower doping level of the GNRs when compared to the conventional polymer-free wet transfer (used for wide nanogaps) and consequently worsen the band-alignment with MoRe. Besides, the GNRs are covered by PMMA after the membrane-transfer, which may by itself influence the doping of the GNRs. 

To get a better understanding of the quality of the MoRe-GNR contact, a comparison with other contact metals is desirable. Activation energies are not as widely reported in the literature as room temperature resistances~\cite{Llinas2017,Martini2019,Richter2020,Braun2021,Hsu2023}. Thus, to compare with other devices, we use the room-temperature resistance per unit of contact width as a benchmark, similar to what is done for 2D materials~\cite{Allain2015}.
The conductance of needle-like MoRe devices with 10-nm nanogaps is around 1~pS for a contact width of around 20~nm, while for the 2~$\upmu$m wide MoRe contacts an average conductance of approximately 1~nS was found. This translates to a conductance per unit width of $0.05~\mathrm{pS}/\mathrm{nm}$ and $0.5~\mathrm{pS}/\mathrm{nm}$, respectively. 

For comparison, our Pd 10-nm nanogaps exhibit a current of 1~nA at 0.1~V, resulting in a conductance of 10~nS over the contact width of around 20~nm, which translates to $500~\mathrm{pS}/\mathrm{nm}$. With the sample contact geometry and PMMA-assisted GNR transfer technique, the reference Pd nanogaps show significantly lower resistances than the MoRe ones. Together with the reduced temperature dependence, this suggests that the band alignment of the Pd work function with the 9-AGNR valence band is better. This could have been partially expected based on the fact that the work function of Pd ($5.12$~ eV~\cite{Michaelson1977}) is larger than the work function of MoRe ($4.6$ - $4.96$~eV~\cite{Berge1974,Liang2006})\textcolor{black}{ (See SI, section 8, for a schematic of the estimated band alignment)}. Another possible explanation for a larger contact resistance for MoRe contacts is the presence of a thin insulating layer on the MoRe surface. Although MoRe alloys are known to have noble-metal-like properties, surface oxidation is yet possible~\cite{Hoekje1994}.

Recently, 9-AGNR devices with Pt wide-nanogap contacts have been studied at room temperature~\cite{Hsu2023}\textcolor{black}{. It was found that} the devices made by PMMA-based GNR transfer
have a larger contact resistance than those made by
polymer-free GNR transfer. This results suggests that it is well possible that the difference observed between the average conductance of the two types of MoRe devices could similarly in part be an effect of the two different transfer methods. For 1-$\upmu$m wide Pt contacts with 50~nm spacing, the reported average conductance for the devices made with PMMA-based GNR transfer and with polymer-free GNR transfer is $1.0~\mathrm{pS}/\mathrm{nm}$ and $100~\mathrm{pS}/\mathrm{nm}$, respectively. 

As a final remark, we note that our MoRe contacts to GNRs show larger contact resistances than expected based on studies on MoRe-CNT junctions, in which resistances smaller than $1$~M$\Omega$ per nanotube were obtained~\cite{Schneider2012,Kaikkonen2020}. A possible explanation for this difference could be that in studies with CNTs, nanotubes were grown or annealed at temperatures higher than $850$~$^\circ$C on top of the MoRe contacts, resulting in molybdenum-carbon end bonds. \textcolor{black}{Since no such annealing step was performed for our GNR devices, we do not expect chemical bonds between the MoRe and the GNRs.} Another possible explanation for the high resistance of our 9-AGNR devices made with MoRe is that the work function of MoRe is too low to achieve a good p-type contact, while it is closer to the work function of graphene ($4.62$ - $4.7$~eV~\cite{Song2012,Rutkov2020}). This explanation is supported by the ambipolar response in the gate dependence of our MoRe devices, which we show in Fig.~S5 and Fig.~S6 in the SI. This suggests that MoRe could still be a good contact metal for GNRs with smaller band-gaps, such as 17-AGNRs~\cite{Yamaguchi2020}, while larger band-gap GNRs are better contacted by high work function metals such as Pd and Pt.

For 9-AGNR devices with superconducting contacts, further advances could be made by doping the 9-AGNRs or modifying the metal/9-AGNR interface by thermal annealing. To reduce contact resistances with MoRe, GNRs with smaller band gaps could also be considered. In addition, the effective work function of superconductors such as MoRe, Nb or NbTiN could be increased, and thus brought into better alignment with valence band of GNRs, by applying a thin coating of Pd or Pt to their surface, e.g. in a hybrid Pd/MoRe heterostructure. 

\section*{Conclusion}

9-atom-wide armchair GNRs were transferred onto pre-fabricated wide and 10-nm MoRe nanogap contacts, as well as onto Pd electrodes with 10-nm nanogaps. To facilitate GNR transfer onto chemically fragile electrodes, we introduce a PMMA-membrane assisted transfer technique for the 10-nm MoRe nanogap contacts, which allows for controlled handling and micron-precise placement of the GNR film without exposing the electrodes to any liquid. We characterized the conductance of the devices as a function of bias voltage and temperature. 
In the devices, the electrical resistance increases with decreasing temperature. The $T$-dependence of the $IV$ characteristics of the wide MoRe nanogap devices can be described by a nuclear tunneling model with a dimensionless dissipation coefficient $\alpha = 9$ and $n = 3$ hopping sites. This is despite the average GNR size of 45~nm exceeding the electrode separation of 30~nm. In comparison, the needle-like MoRe 10-nm nanogaps show a stronger $T$-dependence of conductance with an order of magnitude larger activation energies. The 10-nm MoRe nanogaps also show field-effect response to the local gate, indicating a p-type metal-semiconductor contact. Pd nanogaps show four orders of magnitude higher conductance for the same bias voltage at room temperature with smaller activation energies than MoRe nanogaps. That suggests the possibility of using Pd/MoRe bilayer thin film electrodes for obtaining low-resistance electrical contacts to GNRs for the realization of superconducting metallic electrodes with a nano-scale geometry down to the limit of addressing a single GNR. 



\subsection*{Methods}

\textit{GNR synthesis.} 9-AGNRs were synthesized from 3',6'-diiodo-1,1':2',1"-terphenyl (DITP)~\cite{DiGiovannantonio2018}. Au(111)/mica (Phasis, Switzerland) surface was cleaned in ultrahigh vacuum by two sputtering/annealing cycles: 1~kV Ar+ for 10~min followed by annealing at 470~$^\circ$C for 10 min. In the next step, the precursor monomer DITP was sublimed onto the Au(111) surface from a quartz crucible heated to 70~$^\circ$C, with the substrate held at room temperature. After deposition of approximately one monolayer DITP, the substrate was heated (0.5~K/s) to 200~$^\circ$C for 10 minutes to activate the polymerization reaction, followed by annealing at 400~$^\circ$C (0.5~K/s) for 10 minutes to form the GNRs via cyclodehydrogenation.

\textit{Fabrication of contact electrodes.} For all devices, 285~nm SiO$_2$ on highly-doped p-type Si substrates were cleaned in red fuming nitric acid for at least 5~min. prior to processing. This was followed by a 5 min. O$_2$ plasma step at a flow of 600~sccm and a power of 600~W in a barrel etcher prior to spin coating of e-beam resist. For the wide-nanogap MoRe devices, the e-beam resist AR-P 6200.04 was spin-coated onto the substrates at 4000~RPM and baked at 185~$^{\circ}$C for 3~min., resulting in an 80~nm thick film.

The wide-nanogap geometry was written by an e-beam pattern generator at 100~kV with a $200$~$\upmu$m aperture at a beam current of $800$~pA. The exposed pattern was developed in pentyl acetate for 1~min., followed by a 5~s descum in xylene~\cite{Thoms2014} and a 30~s rinse in isopropyl alcohol, followed by N$_2$ blow drying. A 62:38 MoRe alloy was RF sputtered at an argon pressure of 15~$\upmu$bar and a power of 100~$W$ for 1~min to yield an approximately 20~nm thick layer. Finally, metal lift-off was performed by soaking the samples in the remover AR 600-71 at 70~$^{\circ}$C, followed by a rinse in isopropyl alcohol and N$_2$ blow drying. 

For the e-beam exposure of the needle-like nanogap pattern an overdose-undersize (ODUS) procedure with a shape-proximity error correction algorithm was performed~\cite{Liu2002,Martens2013}. The development was done in pentyl acetate developer that was cooled to $-16$~$^\circ$C in a freezer, as this has been shown to improve contrast for resists that are exposed by polymer chain scission~\cite{Ocola2006}. Upon overdosing to 1000 - 2000 $\upmu\mathrm{C\ cm^{-2}}$ the pattern, 6 - 10~nm electrode separations result. The contact pads were kept further than 100~$\upmu$m from the nanogaps to minimize exposure by back-scattered electrons and exposed at 600~$\upmu\mathrm{C\ cm^{-2}}$. MoRe was sputtered in the same manner as described for the wide-nanogap contacts. 5~nm Ti and 12~nm Pd were deposited by e-beam evaporation at a rate of $0.05$~nm/s and $0.1$~nm/s respectively. 

The MoRe needle-like electrodes were made on top of an evaporated 5~nm Ti + 10~nm Pt local bottom gate covered by a 12~nm thick HfO$_2$ high-$\upkappa$ dielectric layer made by atomic layer deposition at 110~$^\circ$C. 

\textit{PMMA-membrane fabrication}

The procedure for suspension and transfer of the PMMA membranes\textcolor{black}{, based on the technique by Kaverzin et al.\cite{Kaverzin_2022},} is as follows; We start with spin coating a thick layer of a water-soluble polymer, in this case, Elektra~92, onto silicon oxide wafers. Next, a 1000 $\upmu$m thick layer of PMMA~950K was spin-coated onto the Elektra~92 layer in two steps, baked at 180~$^\circ$C for 1 min after each step. A rectangular hole was cut in a piece of scotch tape, which was subsequently pressed against the PMMA membrane on silicon. The scotch tape was suspended over a beaker filled with DI water with the silicon piece submerged to dissolve the Elektra~92. After the silicon piece detached from the PMMA membrane, the membrane was rinsed and left to dry in ambient conditions. The dry PMMA membrane was next stretched over a piece of Polydimethylsiloxane (PDMS), placed on a glass slide.

\section{Acknowledgments}
This work was supported by the Netherlands Organisation for Scientific Research (NWO/OCW), as part of the Frontiers of Nanoscience program. 
T.S.G. also received funding from the European Union Horizon 2020 research and innovation program under grant agreement no. 863098 (SPRING).
G.B.B., P.R. and R.F acknowledge the European Union Horizon 2020 research and innovation program under grant agreement no. 881603 (GrapheneFlagship Core 3) and the Office of Naval Research BRC Program under the grant N00014-18-1-2708. G.B.B., P.R.
and R.F. also greatly appreciate the financial support from the Werner Siemens Foundation (CarboQuant).
This research was funded, in whole or in part by the Swiss National Science Foundation under Grant No. 200020\_182015. For the purpose of Open Access, the author has applied a CC BY public copyright license to any Author Accepted Manuscript version arising from this submission


\section*{Supporting information}
See the supporting information for \textcolor{black}{an} analysis of additional devices\textcolor{black}{, additional electrical characterization, Raman spectroscopy characterization of the transferred 9-AGNR films, band diagrams and optical microscopy images of the MoRe 9-AGNR devices.}

\section*{Author Declarations}
\subsection*{Conflict of Interest}

    The authors have no conflicts to disclose. 

\section*{Data Availability Statement}

The data that support the findings of this study are available from the corresponding author upon request.

\bibliography{Main}

\providecommand{\noopsort}[1]{}\providecommand{\singleletter}[1]{#1}%
\providecommand{\latin}[1]{#1}
\makeatletter
\providecommand{\doi}
  {\begingroup\let\do\@makeother\dospecials
  \catcode`\{=1 \catcode`\}=2 \doi@aux}
\providecommand{\doi@aux}[1]{\endgroup\texttt{#1}}
\makeatother
\providecommand*\mcitethebibliography{\thebibliography}
\csname @ifundefined\endcsname{endmcitethebibliography}
  {\let\endmcitethebibliography\endthebibliography}{}
\begin{mcitethebibliography}{65}
\providecommand*\natexlab[1]{#1}
\providecommand*\mciteSetBstSublistMode[1]{}
\providecommand*\mciteSetBstMaxWidthForm[2]{}
\providecommand*\mciteBstWouldAddEndPuncttrue
  {\def\EndOfBibitem{\unskip.}}
\providecommand*\mciteBstWouldAddEndPunctfalse
  {\let\EndOfBibitem\relax}
\providecommand*\mciteSetBstMidEndSepPunct[3]{}
\providecommand*\mciteSetBstSublistLabelBeginEnd[3]{}
\providecommand*\EndOfBibitem{}
\mciteSetBstSublistMode{f}
\mciteSetBstMaxWidthForm{subitem}{(\alph{mcitesubitemcount})}
\mciteSetBstSublistLabelBeginEnd
  {\mcitemaxwidthsubitemform\space}
  {\relax}
  {\relax}

\bibitem[Nakada \latin{et~al.}(1996)Nakada, Fujita, Dresselhaus, and
  Dresselhaus]{Nakada1996}
Nakada,~K.; Fujita,~M.; Dresselhaus,~G.; Dresselhaus,~M.~S. Edge state in
  graphene ribbons: Nanometer size effect and edge shape dependence.
  \emph{Phys. Rev. B} \textbf{1996}, \emph{54}, 17954--17961\relax
\mciteBstWouldAddEndPuncttrue
\mciteSetBstMidEndSepPunct{\mcitedefaultmidpunct}
{\mcitedefaultendpunct}{\mcitedefaultseppunct}\relax
\EndOfBibitem
\bibitem[Son \latin{et~al.}(2006)Son, Cohen, and Louie]{Son2006_2}
Son,~Y.-W.; Cohen,~M.~L.; Louie,~S.~G. Energy Gaps in Graphene Nanoribbons.
  \emph{Phys. Rev. Lett.} \textbf{2006}, \emph{97}, 216803\relax
\mciteBstWouldAddEndPuncttrue
\mciteSetBstMidEndSepPunct{\mcitedefaultmidpunct}
{\mcitedefaultendpunct}{\mcitedefaultseppunct}\relax
\EndOfBibitem
\bibitem[Yang \latin{et~al.}(2007)Yang, Park, Son, Cohen, and Louie]{Yang2007}
Yang,~L.; Park,~C.-H.; Son,~Y.-W.; Cohen,~M.~L.; Louie,~S.~G. Quasiparticle
  Energies and Band Gaps in Graphene Nanoribbons. \emph{Phys. Rev. Lett.}
  \textbf{2007}, \emph{99}, 186801\relax
\mciteBstWouldAddEndPuncttrue
\mciteSetBstMidEndSepPunct{\mcitedefaultmidpunct}
{\mcitedefaultendpunct}{\mcitedefaultseppunct}\relax
\EndOfBibitem
\bibitem[Schwierz(2010)]{Schwierz2010}
Schwierz,~F. Graphene transistors. \emph{Nat. Nanotechnol.} \textbf{2010},
  \emph{5}, 487--496\relax
\mciteBstWouldAddEndPuncttrue
\mciteSetBstMidEndSepPunct{\mcitedefaultmidpunct}
{\mcitedefaultendpunct}{\mcitedefaultseppunct}\relax
\EndOfBibitem
\bibitem[Llinas \latin{et~al.}(2017)Llinas, Fairbrother, Borin~Barin, Shi, Lee,
  Wu, Yong~Choi, Braganza, Lear, Kau, Choi, Chen, Pedramrazi, Dumslaff, Narita,
  Feng, M{\"u}llen, Fischer, Zettl, Ruffieux, Yablonovitch, Crommie, Fasel, and
  Bokor]{Llinas2017}
Llinas,~J.~P.; Fairbrother,~A.; Borin~Barin,~G.; Shi,~W.; Lee,~K.; Wu,~S.;
  Yong~Choi,~B.; Braganza,~R.; Lear,~J.; Kau,~N.; Choi,~W.; Chen,~C.;
  Pedramrazi,~Z.; Dumslaff,~T.; Narita,~A.; Feng,~X.; M{\"u}llen,~K.;
  Fischer,~F.; Zettl,~A.; Ruffieux,~P.; Yablonovitch,~E.; Crommie,~M.;
  Fasel,~R.; Bokor,~J. Short-channel field-effect transistors with 9-atom and
  13-atom wide graphene nanoribbons. \emph{Nat. Commun.} \textbf{2017},
  \emph{8}, 633\relax
\mciteBstWouldAddEndPuncttrue
\mciteSetBstMidEndSepPunct{\mcitedefaultmidpunct}
{\mcitedefaultendpunct}{\mcitedefaultseppunct}\relax
\EndOfBibitem
\bibitem[Miao \latin{et~al.}(2021)Miao, Wang, Mu, and Wang]{Miao2021}
Miao,~W.; Wang,~L.; Mu,~X.; Wang,~J. The magical photoelectric and
  optoelectronic properties of graphene nanoribbons and their applications.
  \emph{J. Mater. Chem. C} \textbf{2021}, \emph{9}, 13600--13616\relax
\mciteBstWouldAddEndPuncttrue
\mciteSetBstMidEndSepPunct{\mcitedefaultmidpunct}
{\mcitedefaultendpunct}{\mcitedefaultseppunct}\relax
\EndOfBibitem
\bibitem[Son \latin{et~al.}(2006)Son, Cohen, and Louie]{Son2006}
Son,~Y.-W.; Cohen,~M.~L.; Louie,~S.~G. Half-metallic graphene nanoribbons.
  \emph{Nature} \textbf{2006}, \emph{444}, 347--349\relax
\mciteBstWouldAddEndPuncttrue
\mciteSetBstMidEndSepPunct{\mcitedefaultmidpunct}
{\mcitedefaultendpunct}{\mcitedefaultseppunct}\relax
\EndOfBibitem
\bibitem[Tao \latin{et~al.}(2011)Tao, Jiao, Yazyev, Chen, Feng, Zhang, Capaz,
  Tour, Zettl, Louie, Dai, and Crommie]{Tao2011}
Tao,~C.; Jiao,~L.; Yazyev,~O.~V.; Chen,~Y.-C.; Feng,~J.; Zhang,~X.;
  Capaz,~R.~B.; Tour,~J.~M.; Zettl,~A.; Louie,~S.~G.; Dai,~H.; Crommie,~M.~F.
  Spatially resolving edge states of chiral graphene nanoribbons. \emph{Nat.
  Phys.} \textbf{2011}, \emph{7}, 616--620\relax
\mciteBstWouldAddEndPuncttrue
\mciteSetBstMidEndSepPunct{\mcitedefaultmidpunct}
{\mcitedefaultendpunct}{\mcitedefaultseppunct}\relax
\EndOfBibitem
\bibitem[Martin and Blanter(2009)Martin, and Blanter]{Martin2009}
Martin,~I.; Blanter,~Y.~M. Transport in disordered graphene nanoribbons.
  \emph{Phys. Rev. B} \textbf{2009}, \emph{79}, 235132\relax
\mciteBstWouldAddEndPuncttrue
\mciteSetBstMidEndSepPunct{\mcitedefaultmidpunct}
{\mcitedefaultendpunct}{\mcitedefaultseppunct}\relax
\EndOfBibitem
\bibitem[Brede \latin{et~al.}(2023)Brede, Merino-Díez, Berdonces, Sanz,
  Domínguez-Celorrio, Lobo-Checa, Vilas-Varela, Peña, Frederiksen, Pascual,
  de~Oteyza, and Serrate]{Brede2023}
Brede,~J.; Merino-Díez,~N.; Berdonces,~A.; Sanz,~S.; Domínguez-Celorrio,~A.;
  Lobo-Checa,~J.; Vilas-Varela,~M.; Peña,~D.; Frederiksen,~T.; Pascual,~J.~I.;
  de~Oteyza,~D.~G.; Serrate,~D. Detecting the spin-polarization of edge states
  in graphene nanoribbons. 2023; \url{https://arxiv.org/abs/2301.11370}\relax
\mciteBstWouldAddEndPuncttrue
\mciteSetBstMidEndSepPunct{\mcitedefaultmidpunct}
{\mcitedefaultendpunct}{\mcitedefaultseppunct}\relax
\EndOfBibitem
\bibitem[Cai \latin{et~al.}(2010)Cai, Ruffieux, Jaafar, Bieri, Braun,
  Blankenburg, Muoth, Seitsonen, Saleh, Feng, M{\"u}llen, and Fasel]{Cai2010}
Cai,~J.; Ruffieux,~P.; Jaafar,~R.; Bieri,~M.; Braun,~T.; Blankenburg,~S.;
  Muoth,~M.; Seitsonen,~A.~P.; Saleh,~M.; Feng,~X.; M{\"u}llen,~K.; Fasel,~R.
  Atomically precise bottom-up fabrication of graphene nanoribbons.
  \emph{Nature} \textbf{2010}, \emph{466}, 470--473\relax
\mciteBstWouldAddEndPuncttrue
\mciteSetBstMidEndSepPunct{\mcitedefaultmidpunct}
{\mcitedefaultendpunct}{\mcitedefaultseppunct}\relax
\EndOfBibitem
\bibitem[Ruffieux \latin{et~al.}(2016)Ruffieux, Wang, Yang,
  S{\'a}nchez-S{\'a}nchez, Liu, Dienel, Talirz, Shinde, Pignedoli, Passerone,
  Dumslaff, Feng, M{\"u}llen, and Fasel]{Ruffieux2016}
Ruffieux,~P.; Wang,~S.; Yang,~B.; S{\'a}nchez-S{\'a}nchez,~C.; Liu,~J.;
  Dienel,~T.; Talirz,~L.; Shinde,~P.; Pignedoli,~C.~A.; Passerone,~D.;
  Dumslaff,~T.; Feng,~X.; M{\"u}llen,~K.; Fasel,~R. On-surface synthesis of
  graphene nanoribbons with zigzag edge topology. \emph{Nature} \textbf{2016},
  \emph{531}, 489--492\relax
\mciteBstWouldAddEndPuncttrue
\mciteSetBstMidEndSepPunct{\mcitedefaultmidpunct}
{\mcitedefaultendpunct}{\mcitedefaultseppunct}\relax
\EndOfBibitem
\bibitem[Gr{\"o}ning \latin{et~al.}(2018)Gr{\"o}ning, Wang, Yao, Pignedoli,
  Borin~Barin, Daniels, Cupo, Meunier, Feng, Narita, M{\"u}llen, Ruffieux, and
  Fasel]{Groning2018}
Gr{\"o}ning,~O.; Wang,~S.; Yao,~X.; Pignedoli,~C.~A.; Borin~Barin,~G.;
  Daniels,~C.; Cupo,~A.; Meunier,~V.; Feng,~X.; Narita,~A.; M{\"u}llen,~K.;
  Ruffieux,~P.; Fasel,~R. Engineering of robust topological quantum phases in
  graphene nanoribbons. \emph{Nature} \textbf{2018}, \emph{560}, 209--213\relax
\mciteBstWouldAddEndPuncttrue
\mciteSetBstMidEndSepPunct{\mcitedefaultmidpunct}
{\mcitedefaultendpunct}{\mcitedefaultseppunct}\relax
\EndOfBibitem
\bibitem[Rizzo \latin{et~al.}(2018)Rizzo, Veber, Cao, Bronner, Chen, Zhao,
  Rodriguez, Louie, Crommie, and Fischer]{Rizzo2018}
Rizzo,~D.~J.; Veber,~G.; Cao,~T.; Bronner,~C.; Chen,~T.; Zhao,~F.;
  Rodriguez,~H.; Louie,~S.~G.; Crommie,~M.~F.; Fischer,~F.~R. Topological band
  engineering of graphene nanoribbons. \emph{Nature} \textbf{2018}, \emph{560},
  204--208\relax
\mciteBstWouldAddEndPuncttrue
\mciteSetBstMidEndSepPunct{\mcitedefaultmidpunct}
{\mcitedefaultendpunct}{\mcitedefaultseppunct}\relax
\EndOfBibitem
\bibitem[Rizzo \latin{et~al.}(2020)Rizzo, Veber, Jiang, McCurdy, Cao, Bronner,
  Chen, Louie, Fischer, and Crommie]{Rizzo2020}
Rizzo,~D.~J.; Veber,~G.; Jiang,~J.; McCurdy,~R.; Cao,~T.; Bronner,~C.;
  Chen,~T.; Louie,~S.~G.; Fischer,~F.~R.; Crommie,~M.~F. Inducing metallicity
  in graphene nanoribbons via zero-mode superlattices. \emph{Science}
  \textbf{2020}, \emph{369}, 1597--1603\relax
\mciteBstWouldAddEndPuncttrue
\mciteSetBstMidEndSepPunct{\mcitedefaultmidpunct}
{\mcitedefaultendpunct}{\mcitedefaultseppunct}\relax
\EndOfBibitem
\bibitem[Bronner \latin{et~al.}(2013)Bronner, Stremlau, Gille, Brauße, Haase,
  Hecht, and Tegeder]{Bronner2013}
Bronner,~C.; Stremlau,~S.; Gille,~M.; Brauße,~F.; Haase,~A.; Hecht,~S.;
  Tegeder,~P. Aligning the Band Gap of Graphene Nanoribbons by Monomer Doping.
  \emph{Angew. Chem. Int. Ed. Engl.} \textbf{2013}, \emph{52}, 4422--4425\relax
\mciteBstWouldAddEndPuncttrue
\mciteSetBstMidEndSepPunct{\mcitedefaultmidpunct}
{\mcitedefaultendpunct}{\mcitedefaultseppunct}\relax
\EndOfBibitem
\bibitem[Martini \latin{et~al.}(2019)Martini, Chen, Mishra, Barin, Fantuzzi,
  Ruffieux, Fasel, Feng, Narita, Coletti, Müllen, and Candini]{Martini2019}
Martini,~L.; Chen,~Z.; Mishra,~N.; Barin,~G.~B.; Fantuzzi,~P.; Ruffieux,~P.;
  Fasel,~R.; Feng,~X.; Narita,~A.; Coletti,~C.; Müllen,~K.; Candini,~A.
  Structure-dependent electrical properties of graphene nanoribbon devices with
  graphene electrodes. \emph{Carbon} \textbf{2019}, \emph{146}, 36--43\relax
\mciteBstWouldAddEndPuncttrue
\mciteSetBstMidEndSepPunct{\mcitedefaultmidpunct}
{\mcitedefaultendpunct}{\mcitedefaultseppunct}\relax
\EndOfBibitem
\bibitem[El~Abbassi \latin{et~al.}(2020)El~Abbassi, Perrin, Barin, Sangtarash,
  Overbeck, Braun, Lambert, Sun, Prechtl, Narita, Müllen, Ruffieux, Sadeghi,
  Fasel, and Calame]{Elabbassi2020}
El~Abbassi,~M.; Perrin,~M.~L.; Barin,~G.~B.; Sangtarash,~S.; Overbeck,~J.;
  Braun,~O.; Lambert,~C.~J.; Sun,~Q.; Prechtl,~T.; Narita,~A.; Müllen,~K.;
  Ruffieux,~P.; Sadeghi,~H.; Fasel,~R.; Calame,~M. Controlled Quantum Dot
  Formation in Atomically Engineered Graphene Nanoribbon Field-Effect
  Transistors. \emph{ACS Nano} \textbf{2020}, \emph{14}, 5754--5762, PMID:
  32223259\relax
\mciteBstWouldAddEndPuncttrue
\mciteSetBstMidEndSepPunct{\mcitedefaultmidpunct}
{\mcitedefaultendpunct}{\mcitedefaultseppunct}\relax
\EndOfBibitem
\bibitem[Braun \latin{et~al.}(2021)Braun, Overbeck, {El Abbassi}, Käser,
  Furrer, Olziersky, Flasby, {Borin Barin}, Sun, Darawish, Müllen, Ruffieux,
  Fasel, Shorubalko, Perrin, and Calame]{Braun2021}
Braun,~O.; Overbeck,~J.; {El Abbassi},~M.; Käser,~S.; Furrer,~R.;
  Olziersky,~A.; Flasby,~A.; {Borin Barin},~G.; Sun,~Q.; Darawish,~R.;
  Müllen,~K.; Ruffieux,~P.; Fasel,~R.; Shorubalko,~I.; Perrin,~M.~L.;
  Calame,~M. Optimized graphene electrodes for contacting graphene nanoribbons.
  \emph{Carbon} \textbf{2021}, \emph{184}, 331--339\relax
\mciteBstWouldAddEndPuncttrue
\mciteSetBstMidEndSepPunct{\mcitedefaultmidpunct}
{\mcitedefaultendpunct}{\mcitedefaultseppunct}\relax
\EndOfBibitem
\bibitem[Zhang \latin{et~al.}(2023)Zhang, Braun, Barin, Sangtarash, Overbeck,
  Darawish, Stiefel, Furrer, Olziersky, Müllen, Shorubalko, Daaoub, Ruffieux,
  Fasel, Sadeghi, Perrin, and Calame]{Zhang2022_2}
Zhang,~J.; Braun,~O.; Barin,~G.~B.; Sangtarash,~S.; Overbeck,~J.; Darawish,~R.;
  Stiefel,~M.; Furrer,~R.; Olziersky,~A.; Müllen,~K.; Shorubalko,~I.;
  Daaoub,~A. H.~S.; Ruffieux,~P.; Fasel,~R.; Sadeghi,~H.; Perrin,~M.~L.;
  Calame,~M. Tunable Quantum Dots from Atomically Precise Graphene Nanoribbons
  Using a Multi-Gate Architecture. \emph{Adv. Electron. Mater.} \textbf{2023},
  \emph{9}, 2201204\relax
\mciteBstWouldAddEndPuncttrue
\mciteSetBstMidEndSepPunct{\mcitedefaultmidpunct}
{\mcitedefaultendpunct}{\mcitedefaultseppunct}\relax
\EndOfBibitem
\bibitem[Zhang \latin{et~al.}(2022)Zhang, Qian, Barin, Daaoub, Chen, Müllen,
  Sangtarash, Ruffieux, Fasel, Sadeghi, Zhang, Calame, and Perrin]{Zhang2022}
Zhang,~J.; Qian,~L.; Barin,~G.~B.; Daaoub,~A. H.~S.; Chen,~P.; Müllen,~K.;
  Sangtarash,~S.; Ruffieux,~P.; Fasel,~R.; Sadeghi,~H.; Zhang,~J.; Calame,~M.;
  Perrin,~M.~L. Ultimately-scaled electrodes for contacting individual
  atomically-precise graphene nanoribbons. 2022;
  \url{https://arxiv.org/abs/2209.04353}\relax
\mciteBstWouldAddEndPuncttrue
\mciteSetBstMidEndSepPunct{\mcitedefaultmidpunct}
{\mcitedefaultendpunct}{\mcitedefaultseppunct}\relax
\EndOfBibitem
\bibitem[Mutlu \latin{et~al.}(2021)Mutlu, Lin, Barin, Zhang, Pitner, Wang,
  Darawish, Giovannantonio, Wang, Cai, Passlack, Diaz, Narita, Müllen,
  Fischer, Bandaru, Kummel, Ruffieux, Fasel, and Bokor]{Mutlu2021}
Mutlu,~Z.; Lin,~Y.; Barin,~G.~B.; Zhang,~Z.; Pitner,~G.; Wang,~S.;
  Darawish,~R.; Giovannantonio,~M.~D.; Wang,~H.; Cai,~J.; Passlack,~M.;
  Diaz,~C.~H.; Narita,~A.; Müllen,~K.; Fischer,~F.~R.; Bandaru,~P.;
  Kummel,~A.~C.; Ruffieux,~P.; Fasel,~R.; Bokor,~J. Short-Channel Double-Gate
  FETs with Atomically Precise Graphene Nanoribbons. 2021 IEEE International
  Electron Devices Meeting (IEDM). 2021; pp 37.4.1--37.4.4\relax
\mciteBstWouldAddEndPuncttrue
\mciteSetBstMidEndSepPunct{\mcitedefaultmidpunct}
{\mcitedefaultendpunct}{\mcitedefaultseppunct}\relax
\EndOfBibitem
\bibitem[Borin~Barin \latin{et~al.}(2022)Borin~Barin, Sun, Di~Giovannantonio,
  Du, Wang, Llinas, Mutlu, Lin, Wilhelm, Overbeck, Daniels, Lamparski,
  Sahabudeen, Perrin, Urgel, Mishra, Kinikar, Widmer, Stolz, Bommert,
  Pignedoli, Feng, Calame, Müllen, Narita, Meunier, Bokor, Fasel, and
  Ruffieux]{BorinBarin2022}
Borin~Barin,~G.; Sun,~Q.; Di~Giovannantonio,~M.; Du,~C.-Z.; Wang,~X.-Y.;
  Llinas,~J.~P.; Mutlu,~Z.; Lin,~Y.; Wilhelm,~J.; Overbeck,~J.; Daniels,~C.;
  Lamparski,~M.; Sahabudeen,~H.; Perrin,~M.~L.; Urgel,~J.~I.; Mishra,~S.;
  Kinikar,~A.; Widmer,~R.; Stolz,~S.; Bommert,~M.; Pignedoli,~C.; Feng,~X.;
  Calame,~M.; Müllen,~K.; Narita,~A.; Meunier,~V.; Bokor,~J.; Fasel,~R.;
  Ruffieux,~P. Growth Optimization and Device Integration of Narrow-Bandgap
  Graphene Nanoribbons. \emph{Small} \textbf{2022}, \emph{18}, 2202301\relax
\mciteBstWouldAddEndPuncttrue
\mciteSetBstMidEndSepPunct{\mcitedefaultmidpunct}
{\mcitedefaultendpunct}{\mcitedefaultseppunct}\relax
\EndOfBibitem
\bibitem[Lin \latin{et~al.}(2023)Lin, Mutlu, {Borin Barin}, Hong, Llinas,
  Narita, Singh, Müllen, Ruffieux, Fasel, and Bokor]{Lin2023}
Lin,~Y.~C.; Mutlu,~Z.; {Borin Barin},~G.; Hong,~Y.; Llinas,~J.~P.; Narita,~A.;
  Singh,~H.; Müllen,~K.; Ruffieux,~P.; Fasel,~R.; Bokor,~J. Scaling and
  statistics of bottom-up synthesized armchair graphene nanoribbon transistors.
  \emph{Carbon} \textbf{2023}, \emph{205}, 519--526\relax
\mciteBstWouldAddEndPuncttrue
\mciteSetBstMidEndSepPunct{\mcitedefaultmidpunct}
{\mcitedefaultendpunct}{\mcitedefaultseppunct}\relax
\EndOfBibitem
\bibitem[Richter \latin{et~al.}(2020)Richter, Chen, Tries, Prechtl, Narita,
  M{\"u}llen, Asadi, Bonn, and Kl{\"a}ui]{Richter2020}
Richter,~N.; Chen,~Z.; Tries,~A.; Prechtl,~T.; Narita,~A.; M{\"u}llen,~K.;
  Asadi,~K.; Bonn,~M.; Kl{\"a}ui,~M. Charge transport mechanism in networks of
  armchair graphene nanoribbons. \emph{Sci. Rep.} \textbf{2020}, \emph{10},
  1988\relax
\mciteBstWouldAddEndPuncttrue
\mciteSetBstMidEndSepPunct{\mcitedefaultmidpunct}
{\mcitedefaultendpunct}{\mcitedefaultseppunct}\relax
\EndOfBibitem
\bibitem[Chen \latin{et~al.}(2005)Chen, Appenzeller, Knoch, Lin, and
  Avouris]{Chen2005}
Chen,~Z.; Appenzeller,~J.; Knoch,~J.; Lin,~Y.-m.; Avouris,~P. The Role of
  Metal-Nanotube Contact in the Performance of Carbon Nanotube Field-Effect
  Transistors. \emph{Nano. Lett.} \textbf{2005}, \emph{5}, 1497--1502\relax
\mciteBstWouldAddEndPuncttrue
\mciteSetBstMidEndSepPunct{\mcitedefaultmidpunct}
{\mcitedefaultendpunct}{\mcitedefaultseppunct}\relax
\EndOfBibitem
\bibitem[Svensson and Campbell(2011)Svensson, and Campbell]{Svensson2011}
Svensson,~J.; Campbell,~E. E.~B. Schottky barriers in carbon nanotube-metal
  contacts. \emph{J. Appl. Phys.} \textbf{2011}, \emph{110}, 111101\relax
\mciteBstWouldAddEndPuncttrue
\mciteSetBstMidEndSepPunct{\mcitedefaultmidpunct}
{\mcitedefaultendpunct}{\mcitedefaultseppunct}\relax
\EndOfBibitem
\bibitem[Matsuda \latin{et~al.}(2010)Matsuda, Tahir-Kheli, and
  Goddard]{Matsuda2010}
Matsuda,~Y.; Tahir-Kheli,~J.; Goddard,~W. A.~I. Definitive Band Gaps for
  Single-Wall Carbon Nanotubes. \emph{J. Phys. Chem. Lett.} \textbf{2010},
  \emph{1}, 2946--2950\relax
\mciteBstWouldAddEndPuncttrue
\mciteSetBstMidEndSepPunct{\mcitedefaultmidpunct}
{\mcitedefaultendpunct}{\mcitedefaultseppunct}\relax
\EndOfBibitem
\bibitem[Zhang \latin{et~al.}(2000)Zhang, Franklin, Chen, and Dai]{Zhang2000}
Zhang,~Y.; Franklin,~N.~W.; Chen,~R.~J.; Dai,~H. Metal coating on suspended
  carbon nanotubes and its implication to metal–tube interaction. \emph{Chem.
  Phys. Lett.} \textbf{2000}, \emph{331}, 35--41\relax
\mciteBstWouldAddEndPuncttrue
\mciteSetBstMidEndSepPunct{\mcitedefaultmidpunct}
{\mcitedefaultendpunct}{\mcitedefaultseppunct}\relax
\EndOfBibitem
\bibitem[Sarkar \latin{et~al.}(2014)Sarkar, Moser, Tian, Zhang, Al-Hadeethi,
  and Haddon]{Sarkar2014}
Sarkar,~S.; Moser,~M.~L.; Tian,~X.; Zhang,~X.; Al-Hadeethi,~Y.~F.;
  Haddon,~R.~C. Metals on Graphene and Carbon Nanotube Surfaces: From Mobile
  Atoms to Atomtronics to Bulk Metals to Clusters and Catalysts. \emph{Chem.
  Mater.} \textbf{2014}, \emph{26}, 184--195\relax
\mciteBstWouldAddEndPuncttrue
\mciteSetBstMidEndSepPunct{\mcitedefaultmidpunct}
{\mcitedefaultendpunct}{\mcitedefaultseppunct}\relax
\EndOfBibitem
\bibitem[Javey \latin{et~al.}(2003)Javey, Guo, Wang, Lundstrom, and
  Dai]{Javey2003}
Javey,~A.; Guo,~J.; Wang,~Q.; Lundstrom,~M.; Dai,~H. Ballistic carbon nanotube
  field-effect transistors. \emph{Nature} \textbf{2003}, \emph{424},
  654--657\relax
\mciteBstWouldAddEndPuncttrue
\mciteSetBstMidEndSepPunct{\mcitedefaultmidpunct}
{\mcitedefaultendpunct}{\mcitedefaultseppunct}\relax
\EndOfBibitem
\bibitem[Mann \latin{et~al.}(2003)Mann, Javey, Kong, Wang, and Dai]{Mann2003}
Mann,~D.; Javey,~A.; Kong,~J.; Wang,~Q.; Dai,~H. Ballistic Transport in
  Metallic Nanotubes with Reliable Pd Ohmic Contacts. \emph{Nano. Lett.}
  \textbf{2003}, \emph{3}, 1541--1544\relax
\mciteBstWouldAddEndPuncttrue
\mciteSetBstMidEndSepPunct{\mcitedefaultmidpunct}
{\mcitedefaultendpunct}{\mcitedefaultseppunct}\relax
\EndOfBibitem
\bibitem[Cao \latin{et~al.}(2015)Cao, Han, Tersoff, Zhu, Zhang, Tulevski, Tang,
  and Haensch]{Cao2015}
Cao,~Q.; Han,~S.~J.; Tersoff,~J.; Zhu,~Y.; Zhang,~Z.; Tulevski,~G.~S.;
  Tang,~J.; Haensch,~W. End-bonded contacts for carbon nanotube transistors
  with low, size-independent resistance. \emph{Science} \textbf{2015},
  \emph{350}, 68--72\relax
\mciteBstWouldAddEndPuncttrue
\mciteSetBstMidEndSepPunct{\mcitedefaultmidpunct}
{\mcitedefaultendpunct}{\mcitedefaultseppunct}\relax
\EndOfBibitem
\bibitem[Schneider \latin{et~al.}(2012)Schneider, Etaki, van~der Zant, and
  Steele]{Schneider2012}
Schneider,~B.~H.; Etaki,~S.; van~der Zant,~H. S.~J.; Steele,~G.~A. Coupling
  carbon nanotube mechanics to a superconducting circuit. \emph{Sci. Rep.}
  \textbf{2012}, \emph{2}, 599\relax
\mciteBstWouldAddEndPuncttrue
\mciteSetBstMidEndSepPunct{\mcitedefaultmidpunct}
{\mcitedefaultendpunct}{\mcitedefaultseppunct}\relax
\EndOfBibitem
\bibitem[Kaikkonen \latin{et~al.}(2020)Kaikkonen, Sebastian, Laiho, Wei, Will,
  Liao, Kauppinen, and Hakonen]{Kaikkonen2020}
Kaikkonen,~J.-P.; Sebastian,~A.~T.; Laiho,~P.; Wei,~N.; Will,~M.; Liao,~Y.;
  Kauppinen,~E.~I.; Hakonen,~P.~J. Suspended superconducting weak links from
  aerosol-synthesized single-walled carbon nanotubes. \emph{Nano Res.}
  \textbf{2020}, \emph{13}, 3433--3438\relax
\mciteBstWouldAddEndPuncttrue
\mciteSetBstMidEndSepPunct{\mcitedefaultmidpunct}
{\mcitedefaultendpunct}{\mcitedefaultseppunct}\relax
\EndOfBibitem
\bibitem[Meissner(1960)]{Meissner1960}
Meissner,~H. Superconductivity of Contacts with Interposed Barriers.
  \emph{Phys. Rev.} \textbf{1960}, \emph{117}, 672--680\relax
\mciteBstWouldAddEndPuncttrue
\mciteSetBstMidEndSepPunct{\mcitedefaultmidpunct}
{\mcitedefaultendpunct}{\mcitedefaultseppunct}\relax
\EndOfBibitem
\bibitem[Di~Giovannantonio \latin{et~al.}(2018)Di~Giovannantonio, Deniz, Urgel,
  Widmer, Dienel, Stolz, Sánchez-Sánchez, Muntwiler, Dumslaff, Berger,
  Narita, Feng, Müllen, Ruffieux, and Fasel]{DiGiovannantonio2018}
Di~Giovannantonio,~M.; Deniz,~O.; Urgel,~J.~I.; Widmer,~R.; Dienel,~T.;
  Stolz,~S.; Sánchez-Sánchez,~C.; Muntwiler,~M.; Dumslaff,~T.; Berger,~R.;
  Narita,~A.; Feng,~X.; Müllen,~K.; Ruffieux,~P.; Fasel,~R. On-Surface Growth
  Dynamics of Graphene Nanoribbons: The Role of Halogen Functionalization.
  \emph{ACS Nano} \textbf{2018}, \emph{12}, 74--81, PMID: 29200262\relax
\mciteBstWouldAddEndPuncttrue
\mciteSetBstMidEndSepPunct{\mcitedefaultmidpunct}
{\mcitedefaultendpunct}{\mcitedefaultseppunct}\relax
\EndOfBibitem
\bibitem[Borin~Barin \latin{et~al.}(2019)Borin~Barin, Fairbrother, Rotach,
  Bayle, Paillet, Liang, Meunier, Hauert, Dumslaff, Narita, M{\"u}llen,
  Sahabudeen, Berger, Feng, Fasel, and Ruffieux]{BorinBarin2019}
Borin~Barin,~G.; Fairbrother,~A.; Rotach,~L.; Bayle,~M.; Paillet,~M.;
  Liang,~L.; Meunier,~V.; Hauert,~R.; Dumslaff,~T.; Narita,~A.; M{\"u}llen,~K.;
  Sahabudeen,~H.; Berger,~R.; Feng,~X.; Fasel,~R.; Ruffieux,~P.
  Surface-Synthesized Graphene Nanoribbons for Room Temperature Switching
  Devices: Substrate Transfer and ex Situ Characterization. \emph{ACS Appl.
  Nano. Mater.} \textbf{2019}, \emph{2}, 2184--2192\relax
\mciteBstWouldAddEndPuncttrue
\mciteSetBstMidEndSepPunct{\mcitedefaultmidpunct}
{\mcitedefaultendpunct}{\mcitedefaultseppunct}\relax
\EndOfBibitem
\bibitem[Fisher and Dorsey(1985)Fisher, and Dorsey]{Fisher1985}
Fisher,~M. P.~A.; Dorsey,~A.~T. Dissipative Quantum Tunneling in a Biased
  Double-Well System at Finite Temperatures. \emph{Phys. Rev. Lett.}
  \textbf{1985}, \emph{54}, 1609--1612\relax
\mciteBstWouldAddEndPuncttrue
\mciteSetBstMidEndSepPunct{\mcitedefaultmidpunct}
{\mcitedefaultendpunct}{\mcitedefaultseppunct}\relax
\EndOfBibitem
\bibitem[Grabert and Weiss(1985)Grabert, and Weiss]{Grabert1985}
Grabert,~H.; Weiss,~U. Quantum Tunneling Rates for Asymmetric Double-Well
  Systems with Ohmic Dissipation. \emph{Phys. Rev. Lett.} \textbf{1985},
  \emph{54}, 1605--1608\relax
\mciteBstWouldAddEndPuncttrue
\mciteSetBstMidEndSepPunct{\mcitedefaultmidpunct}
{\mcitedefaultendpunct}{\mcitedefaultseppunct}\relax
\EndOfBibitem
\bibitem[Bockrath \latin{et~al.}(1999)Bockrath, Cobden, Lu, Rinzler, Smalley,
  Balents, and McEuen]{Bockrath1999}
Bockrath,~M.; Cobden,~D.~H.; Lu,~J.; Rinzler,~A.~G.; Smalley,~R.~E.;
  Balents,~L.; McEuen,~P.~L. Luttinger-liquid behaviour in carbon nanotubes.
  \emph{Nature} \textbf{1999}, \emph{397}, 598--601\relax
\mciteBstWouldAddEndPuncttrue
\mciteSetBstMidEndSepPunct{\mcitedefaultmidpunct}
{\mcitedefaultendpunct}{\mcitedefaultseppunct}\relax
\EndOfBibitem
\bibitem[Bennett \latin{et~al.}(2013)Bennett, Pedramrazi, Madani, Chen,
  de~Oteyza, Chen, Fischer, Crommie, and Bokor]{Bennett2013}
Bennett,~P.~B.; Pedramrazi,~Z.; Madani,~A.; Chen,~Y.-C.; de~Oteyza,~D.~G.;
  Chen,~C.; Fischer,~F.~R.; Crommie,~M.~F.; Bokor,~J. Bottom-up graphene
  nanoribbon field-effect transistors. \emph{Appl. Phys. Lett.} \textbf{2013},
  \emph{103}, 253114\relax
\mciteBstWouldAddEndPuncttrue
\mciteSetBstMidEndSepPunct{\mcitedefaultmidpunct}
{\mcitedefaultendpunct}{\mcitedefaultseppunct}\relax
\EndOfBibitem
\bibitem[Tries \latin{et~al.}(2020)Tries, Richter, Chen, Narita, Müllen, Wang,
  Bonn, and Kläui]{Tries2020}
Tries,~A.; Richter,~N.; Chen,~Z.; Narita,~A.; Müllen,~K.; Wang,~H.~I.;
  Bonn,~M.; Kläui,~M. Hysteresis in graphene nanoribbon field-effect devices.
  \emph{Phys. Chem. Chem. Phys.} \textbf{2020}, \emph{22}, 5667--5672\relax
\mciteBstWouldAddEndPuncttrue
\mciteSetBstMidEndSepPunct{\mcitedefaultmidpunct}
{\mcitedefaultendpunct}{\mcitedefaultseppunct}\relax
\EndOfBibitem
\bibitem[Overbeck \latin{et~al.}(2019)Overbeck, Borin~Barin, Daniels, Perrin,
  Liang, Braun, Darawish, Burkhardt, Dumslaff, Wang, Narita, Müllen, Meunier,
  Fasel, Calame, and Ruffieux]{Overbeck2019}
Overbeck,~J.; Borin~Barin,~G.; Daniels,~C.; Perrin,~M.~L.; Liang,~L.;
  Braun,~O.; Darawish,~R.; Burkhardt,~B.; Dumslaff,~T.; Wang,~X.-Y.;
  Narita,~A.; Müllen,~K.; Meunier,~V.; Fasel,~R.; Calame,~M.; Ruffieux,~P.
  Optimized Substrates and Measurement Approaches for Raman Spectroscopy of
  Graphene Nanoribbons. \emph{Phys. Status Solidi B} \textbf{2019}, \emph{256},
  1900343\relax
\mciteBstWouldAddEndPuncttrue
\mciteSetBstMidEndSepPunct{\mcitedefaultmidpunct}
{\mcitedefaultendpunct}{\mcitedefaultseppunct}\relax
\EndOfBibitem
\bibitem[Zhang \latin{et~al.}(2009)Zhang, Brar, Girit, Zettl, and
  Crommie]{Zhang2009}
Zhang,~Y.; Brar,~V.~W.; Girit,~C.; Zettl,~A.; Crommie,~M.~F. Origin of spatial
  charge inhomogeneity in graphene. \emph{Nat. Phys.} \textbf{2009}, \emph{5},
  722--726\relax
\mciteBstWouldAddEndPuncttrue
\mciteSetBstMidEndSepPunct{\mcitedefaultmidpunct}
{\mcitedefaultendpunct}{\mcitedefaultseppunct}\relax
\EndOfBibitem
\bibitem[Mayamei \latin{et~al.}(2020)Mayamei, Shin, Watanabe, Taniguchi, and
  Bae]{Mayamei2020}
Mayamei,~Y.; Shin,~J.~C.; Watanabe,~K.; Taniguchi,~T.; Bae,~M.-H. Landscape of
  Charge Puddles in Graphene Nanoribbons on Hexagonal Boron Nitride.
  \emph{Phys. Status Solidi B} \textbf{2020}, \emph{257}, 2000317\relax
\mciteBstWouldAddEndPuncttrue
\mciteSetBstMidEndSepPunct{\mcitedefaultmidpunct}
{\mcitedefaultendpunct}{\mcitedefaultseppunct}\relax
\EndOfBibitem
\bibitem[Fediai \latin{et~al.}(2016)Fediai, Ryndyk, Seifert, Mothes, Claus,
  Schröter, and Cuniberti]{Fediai2016}
Fediai,~A.; Ryndyk,~D.~A.; Seifert,~G.; Mothes,~S.; Claus,~M.; Schröter,~M.;
  Cuniberti,~G. Towards an optimal contact metal for CNTFETs. \emph{Nanoscale}
  \textbf{2016}, \emph{8}, 10240--10251\relax
\mciteBstWouldAddEndPuncttrue
\mciteSetBstMidEndSepPunct{\mcitedefaultmidpunct}
{\mcitedefaultendpunct}{\mcitedefaultseppunct}\relax
\EndOfBibitem
\bibitem[Sutter \latin{et~al.}(2009)Sutter, Sadowski, and Sutter]{Sutter2009}
Sutter,~P.; Sadowski,~J.~T.; Sutter,~E. Graphene on Pt(111): Growth and
  substrate interaction. \emph{Phys. Rev. B} \textbf{2009}, \emph{80},
  245411\relax
\mciteBstWouldAddEndPuncttrue
\mciteSetBstMidEndSepPunct{\mcitedefaultmidpunct}
{\mcitedefaultendpunct}{\mcitedefaultseppunct}\relax
\EndOfBibitem
\bibitem[Yao \latin{et~al.}(2019)Yao, ang Peng, nan Huang, yong Zhang, yuan
  Shi, and Jin]{Yao2019}
Yao,~Y.; ang Peng,~S.; nan Huang,~X.; yong Zhang,~D.; yuan Shi,~J.; Jin,~Z. A
  uniform stable P-type graphene doping method with a gold etching process.
  \emph{Nanotechnology} \textbf{2019}, \emph{30}, 405205\relax
\mciteBstWouldAddEndPuncttrue
\mciteSetBstMidEndSepPunct{\mcitedefaultmidpunct}
{\mcitedefaultendpunct}{\mcitedefaultseppunct}\relax
\EndOfBibitem
\bibitem[Wu \latin{et~al.}(2012)Wu, Jiang, Ren, Cai, Lee, Li, Piner, Pope, Hao,
  Ji, Kang, and Ruoff]{Wu2012}
Wu,~Y.; Jiang,~W.; Ren,~Y.; Cai,~W.; Lee,~W.~H.; Li,~H.; Piner,~R.~D.;
  Pope,~C.~W.; Hao,~Y.; Ji,~H.; Kang,~J.; Ruoff,~R.~S. Tuning the Doping Type
  and Level of Graphene with Different Gold Configurations. \emph{Small}
  \textbf{2012}, \emph{8}, 3129--3136\relax
\mciteBstWouldAddEndPuncttrue
\mciteSetBstMidEndSepPunct{\mcitedefaultmidpunct}
{\mcitedefaultendpunct}{\mcitedefaultseppunct}\relax
\EndOfBibitem
\bibitem[Hsu \latin{et~al.}(2023)Hsu, Rohde, Borin~Barin, Gandus, Passerone,
  Luisier, Ruffieux, Fasel, van~der Zant, and Abbassi]{Hsu2023}
Hsu,~C.; Rohde,~M.; Borin~Barin,~G.; Gandus,~G.; Passerone,~D.; Luisier,~M.;
  Ruffieux,~P.; Fasel,~R.; van~der Zant,~H. S.~J.; Abbassi,~M.~E. {Platinum
  contacts for 9-atom-wide armchair graphene nanoribbons}. \emph{Appl. Phys.
  Lett.} \textbf{2023}, \emph{122}, 173104\relax
\mciteBstWouldAddEndPuncttrue
\mciteSetBstMidEndSepPunct{\mcitedefaultmidpunct}
{\mcitedefaultendpunct}{\mcitedefaultseppunct}\relax
\EndOfBibitem
\bibitem[Allain \latin{et~al.}(2015)Allain, Kang, Banerjee, and
  Kis]{Allain2015}
Allain,~A.; Kang,~J.; Banerjee,~K.; Kis,~A. Electrical contacts to
  two-dimensional semiconductors. \emph{Nat. Mater.} \textbf{2015}, \emph{14},
  1195--1205\relax
\mciteBstWouldAddEndPuncttrue
\mciteSetBstMidEndSepPunct{\mcitedefaultmidpunct}
{\mcitedefaultendpunct}{\mcitedefaultseppunct}\relax
\EndOfBibitem
\bibitem[Michaelson(1977)]{Michaelson1977}
Michaelson,~H.~B. The work function of the elements and its periodicity.
  \emph{J. Appl. Phys.} \textbf{1977}, \emph{48}, 4729--4733\relax
\mciteBstWouldAddEndPuncttrue
\mciteSetBstMidEndSepPunct{\mcitedefaultmidpunct}
{\mcitedefaultendpunct}{\mcitedefaultseppunct}\relax
\EndOfBibitem
\bibitem[Berge \latin{et~al.}(1974)Berge, Gartland, and Slagsvold]{Berge1974}
Berge,~S.; Gartland,~P.; Slagsvold,~B. Photoelectric work function of a
  molybdenum single crystal for the (100), (110), (111), (112), (114), and
  (332) faces. \emph{Surf. Sci.} \textbf{1974}, \emph{43}, 275--292\relax
\mciteBstWouldAddEndPuncttrue
\mciteSetBstMidEndSepPunct{\mcitedefaultmidpunct}
{\mcitedefaultendpunct}{\mcitedefaultseppunct}\relax
\EndOfBibitem
\bibitem[Liang \latin{et~al.}(2006)Liang, Curless, Tracy, Gilmer, Schaeffer,
  Triyoso, and Tobin]{Liang2006}
Liang,~Y.; Curless,~J.; Tracy,~C.~J.; Gilmer,~D.~C.; Schaeffer,~J.~K.;
  Triyoso,~D.~H.; Tobin,~P.~J. Interface dipole and effective work function of
  Re in Re/HfO2/SiOx/n-Si gate stack. \emph{Appl. Phys. Lett.} \textbf{2006},
  \emph{88}, 072907\relax
\mciteBstWouldAddEndPuncttrue
\mciteSetBstMidEndSepPunct{\mcitedefaultmidpunct}
{\mcitedefaultendpunct}{\mcitedefaultseppunct}\relax
\EndOfBibitem
\bibitem[Hoekje \latin{et~al.}(1993)Hoekje, Outlaw, and Sankaran]{Hoekje1994}
Hoekje,~S.; Outlaw,~R.; Sankaran,~S. Surface compositional variations of
  Mo-47Re alloy as a function of temperature. \emph{NASA Tech. Pap. 3402}
  \textbf{1993}, \relax
\mciteBstWouldAddEndPunctfalse
\mciteSetBstMidEndSepPunct{\mcitedefaultmidpunct}
{}{\mcitedefaultseppunct}\relax
\EndOfBibitem
\bibitem[Song \latin{et~al.}(2012)Song, Park, Sul, and Cho]{Song2012}
Song,~S.~M.; Park,~J.~K.; Sul,~O.~J.; Cho,~B.~J. Determination of Work Function
  of Graphene under a Metal Electrode and Its Role in Contact Resistance.
  \emph{Nano. Lett.} \textbf{2012}, \emph{12}, 3887--3892\relax
\mciteBstWouldAddEndPuncttrue
\mciteSetBstMidEndSepPunct{\mcitedefaultmidpunct}
{\mcitedefaultendpunct}{\mcitedefaultseppunct}\relax
\EndOfBibitem
\bibitem[Rut'kov \latin{et~al.}(2020)Rut'kov, Afanas'eva, and Gall]{Rutkov2020}
Rut'kov,~E.; Afanas'eva,~E.; Gall,~N. Graphene and graphite work function
  depending on layer number on Re. \emph{Diam. Relat. Mater.} \textbf{2020},
  \emph{101}, 107576\relax
\mciteBstWouldAddEndPuncttrue
\mciteSetBstMidEndSepPunct{\mcitedefaultmidpunct}
{\mcitedefaultendpunct}{\mcitedefaultseppunct}\relax
\EndOfBibitem
\bibitem[Yamaguchi \latin{et~al.}(2020)Yamaguchi, Hayashi, Jippo, Shiotari,
  Ohtomo, Sakakura, Hieda, Aratani, Ohfuchi, Sugimoto, Yamada, and
  Sato]{Yamaguchi2020}
Yamaguchi,~J.; Hayashi,~H.; Jippo,~H.; Shiotari,~A.; Ohtomo,~M.; Sakakura,~M.;
  Hieda,~N.; Aratani,~N.; Ohfuchi,~M.; Sugimoto,~Y.; Yamada,~H.; Sato,~S. Small
  bandgap in atomically precise 17-atom-wide armchair-edged graphene
  nanoribbons. \emph{Commun. Mater.} \textbf{2020}, \emph{1}, 36\relax
\mciteBstWouldAddEndPuncttrue
\mciteSetBstMidEndSepPunct{\mcitedefaultmidpunct}
{\mcitedefaultendpunct}{\mcitedefaultseppunct}\relax
\EndOfBibitem
\bibitem[Thoms and Macintyre(2014)Thoms, and Macintyre]{Thoms2014}
Thoms,~S.; Macintyre,~D.~S. Investigation of CSAR 62, a new resist for electron
  beam lithography. \emph{J. Vac. Sci. Technol. B} \textbf{2014}, \emph{32},
  06FJ01\relax
\mciteBstWouldAddEndPuncttrue
\mciteSetBstMidEndSepPunct{\mcitedefaultmidpunct}
{\mcitedefaultendpunct}{\mcitedefaultseppunct}\relax
\EndOfBibitem
\bibitem[Liu \latin{et~al.}(2002)Liu, Avouris, Bucchignano, Martel, Sun, and
  Michl]{Liu2002}
Liu,~K.; Avouris,~P.; Bucchignano,~J.; Martel,~R.; Sun,~S.; Michl,~J. Simple
  fabrication scheme for sub-10 nm electrode gaps using electron-beam
  lithography. \emph{Appl. Phys. Lett.} \textbf{2002}, \emph{80},
  865--867\relax
\mciteBstWouldAddEndPuncttrue
\mciteSetBstMidEndSepPunct{\mcitedefaultmidpunct}
{\mcitedefaultendpunct}{\mcitedefaultseppunct}\relax
\EndOfBibitem
\bibitem[Martens \latin{et~al.}(2013)Martens, Butschke, Galler, Kr{\"u}ger,
  Sailer, and S{\"u}lzle]{Martens2013}
Martens,~S.; Butschke,~J.; Galler,~R.; Kr{\"u}ger,~M.; Sailer,~H.;
  S{\"u}lzle,~M. {E-beam GIDC resolution enhancement technology in practical
  applications}. Photomask Technology 2013. 2013; p 88802H\relax
\mciteBstWouldAddEndPuncttrue
\mciteSetBstMidEndSepPunct{\mcitedefaultmidpunct}
{\mcitedefaultendpunct}{\mcitedefaultseppunct}\relax
\EndOfBibitem
\bibitem[Ocola and Stein(2006)Ocola, and Stein]{Ocola2006}
Ocola,~L.~E.; Stein,~A. Effect of cold development on improvement in
  electron-beam nanopatterning resolution and line roughness. \emph{J. Vac.
  Sci. Technol. B Nanotechnol. Microelectron.} \textbf{2006}, \emph{24},
  3061--3065\relax
\mciteBstWouldAddEndPuncttrue
\mciteSetBstMidEndSepPunct{\mcitedefaultmidpunct}
{\mcitedefaultendpunct}{\mcitedefaultseppunct}\relax
\EndOfBibitem
\bibitem[Kaverzin \latin{et~al.}(2022)Kaverzin, Ghiasi, Dismukes, Roy, and van
  Wees]{Kaverzin_2022}
Kaverzin,~A.~A.; Ghiasi,~T.~S.; Dismukes,~A.~H.; Roy,~X.; van Wees,~B.~J. Spin
  injection by spin–charge coupling in proximity induced magnetic graphene.
  \emph{2D Mater.} \textbf{2022}, \emph{9}, 045003\relax
\mciteBstWouldAddEndPuncttrue
\mciteSetBstMidEndSepPunct{\mcitedefaultmidpunct}
{\mcitedefaultendpunct}{\mcitedefaultseppunct}\relax
\EndOfBibitem
\end{mcitethebibliography}


\providecommand{\noopsort}[1]{}\providecommand{\singleletter}[1]{#1}%
\providecommand{\latin}[1]{#1}
\makeatletter
\providecommand{\doi}
  {\begingroup\let\do\@makeother\dospecials
  \catcode`\{=1 \catcode`\}=2 \doi@aux}
\providecommand{\doi@aux}[1]{\endgroup\texttt{#1}}
\makeatother
\providecommand*\mcitethebibliography{\thebibliography}
\csname @ifundefined\endcsname{endmcitethebibliography}
  {\let\endmcitethebibliography\endthebibliography}{}
\begin{mcitethebibliography}{17}
\providecommand*\natexlab[1]{#1}
\providecommand*\mciteSetBstSublistMode[1]{}
\providecommand*\mciteSetBstMaxWidthForm[2]{}
\providecommand*\mciteBstWouldAddEndPuncttrue
  {\def\EndOfBibitem{\unskip.}}
\providecommand*\mciteBstWouldAddEndPunctfalse
  {\let\EndOfBibitem\relax}
\providecommand*\mciteSetBstMidEndSepPunct[3]{}
\providecommand*\mciteSetBstSublistLabelBeginEnd[3]{}
\providecommand*\EndOfBibitem{}
\mciteSetBstSublistMode{f}
\mciteSetBstMaxWidthForm{subitem}{(\alph{mcitesubitemcount})}
\mciteSetBstSublistLabelBeginEnd
  {\mcitemaxwidthsubitemform\space}
  {\relax}
  {\relax}

\bibitem[Wang \latin{et~al.}(2010)Wang, Wu, Cong, Shang, and Yu]{Wang2010}
Wang,~H.; Wu,~Y.; Cong,~C.; Shang,~J.; Yu,~T. Hysteresis of Electronic
  Transport in Graphene Transistors. \emph{ACS Nano} \textbf{2010}, \emph{4},
  7221--7228\relax
\mciteBstWouldAddEndPuncttrue
\mciteSetBstMidEndSepPunct{\mcitedefaultmidpunct}
{\mcitedefaultendpunct}{\mcitedefaultseppunct}\relax
\EndOfBibitem
\bibitem[Kim \latin{et~al.}(2003)Kim, Javey, Vermesh, Wang, Li, and
  Dai]{Kim2003}
Kim,~W.; Javey,~A.; Vermesh,~O.; Wang,~Q.; Li,~Y.; Dai,~H. Hysteresis Caused by
  Water Molecules in Carbon Nanotube Field-Effect Transistors. \emph{Nano.
  Lett.} \textbf{2003}, \emph{3}, 193--198\relax
\mciteBstWouldAddEndPuncttrue
\mciteSetBstMidEndSepPunct{\mcitedefaultmidpunct}
{\mcitedefaultendpunct}{\mcitedefaultseppunct}\relax
\EndOfBibitem
\bibitem[Xu \latin{et~al.}(2012)Xu, Chen, Zhang, and Zhang]{Xu2012}
Xu,~H.; Chen,~Y.; Zhang,~J.; Zhang,~H. Investigating the Mechanism of
  Hysteresis Effect in Graphene Electrical Field Device Fabricated on SiO2
  Substrates using Raman Spectroscopy. \emph{Small} \textbf{2012}, \emph{8},
  2833--2840\relax
\mciteBstWouldAddEndPuncttrue
\mciteSetBstMidEndSepPunct{\mcitedefaultmidpunct}
{\mcitedefaultendpunct}{\mcitedefaultseppunct}\relax
\EndOfBibitem
\bibitem[Tries \latin{et~al.}(2020)Tries, Richter, Chen, Narita, Müllen, Wang,
  Bonn, and Kläui]{Tries2020}
Tries,~A.; Richter,~N.; Chen,~Z.; Narita,~A.; Müllen,~K.; Wang,~H.~I.;
  Bonn,~M.; Kläui,~M. Hysteresis in graphene nanoribbon field-effect devices.
  \emph{Phys. Chem. Chem. Phys.} \textbf{2020}, \emph{22}, 5667--5672\relax
\mciteBstWouldAddEndPuncttrue
\mciteSetBstMidEndSepPunct{\mcitedefaultmidpunct}
{\mcitedefaultendpunct}{\mcitedefaultseppunct}\relax
\EndOfBibitem
\bibitem[Bennett \latin{et~al.}(2013)Bennett, Pedramrazi, Madani, Chen,
  de~Oteyza, Chen, Fischer, Crommie, and Bokor]{Bennett2013}
Bennett,~P.~B.; Pedramrazi,~Z.; Madani,~A.; Chen,~Y.-C.; de~Oteyza,~D.~G.;
  Chen,~C.; Fischer,~F.~R.; Crommie,~M.~F.; Bokor,~J. Bottom-up graphene
  nanoribbon field-effect transistors. \emph{Appl. Phys. Lett.} \textbf{2013},
  \emph{103}, 253114\relax
\mciteBstWouldAddEndPuncttrue
\mciteSetBstMidEndSepPunct{\mcitedefaultmidpunct}
{\mcitedefaultendpunct}{\mcitedefaultseppunct}\relax
\EndOfBibitem
\bibitem[Lu \latin{et~al.}(2022)Lu, Lin, Tsai, and Lin]{Lu2022}
Lu,~Y.-X.; Lin,~C.-T.; Tsai,~M.-H.; Lin,~K.-C. Review-Hysteresis in Carbon
  Nano-Structure Field Effect Transistor. \emph{Micromachines} \textbf{2022},
  \emph{13}\relax
\mciteBstWouldAddEndPuncttrue
\mciteSetBstMidEndSepPunct{\mcitedefaultmidpunct}
{\mcitedefaultendpunct}{\mcitedefaultseppunct}\relax
\EndOfBibitem
\bibitem[Overbeck \latin{et~al.}(2019)Overbeck, Borin~Barin, Daniels, Perrin,
  Liang, Braun, Darawish, Burkhardt, Dumslaff, Wang, Narita, Müllen, Meunier,
  Fasel, Calame, and Ruffieux]{Overbeck2019}
Overbeck,~J.; Borin~Barin,~G.; Daniels,~C.; Perrin,~M.~L.; Liang,~L.;
  Braun,~O.; Darawish,~R.; Burkhardt,~B.; Dumslaff,~T.; Wang,~X.-Y.;
  Narita,~A.; Müllen,~K.; Meunier,~V.; Fasel,~R.; Calame,~M.; Ruffieux,~P.
  Optimized Substrates and Measurement Approaches for Raman Spectroscopy of
  Graphene Nanoribbons. \emph{Phys. Status Solidi B} \textbf{2019}, \emph{256},
  1900343\relax
\mciteBstWouldAddEndPuncttrue
\mciteSetBstMidEndSepPunct{\mcitedefaultmidpunct}
{\mcitedefaultendpunct}{\mcitedefaultseppunct}\relax
\EndOfBibitem
\bibitem[Borin~Barin \latin{et~al.}(2019)Borin~Barin, Fairbrother, Rotach,
  Bayle, Paillet, Liang, Meunier, Hauert, Dumslaff, Narita, M{\"u}llen,
  Sahabudeen, Berger, Feng, Fasel, and Ruffieux]{BorinBarin2019}
Borin~Barin,~G.; Fairbrother,~A.; Rotach,~L.; Bayle,~M.; Paillet,~M.;
  Liang,~L.; Meunier,~V.; Hauert,~R.; Dumslaff,~T.; Narita,~A.; M{\"u}llen,~K.;
  Sahabudeen,~H.; Berger,~R.; Feng,~X.; Fasel,~R.; Ruffieux,~P.
  Surface-Synthesized Graphene Nanoribbons for Room Temperature Switching
  Devices: Substrate Transfer and ex Situ Characterization. \emph{ACS Appl.
  Nano. Mater.} \textbf{2019}, \emph{2}, 2184--2192\relax
\mciteBstWouldAddEndPuncttrue
\mciteSetBstMidEndSepPunct{\mcitedefaultmidpunct}
{\mcitedefaultendpunct}{\mcitedefaultseppunct}\relax
\EndOfBibitem
\bibitem[Lee and Chang(2018)Lee, and Chang]{Lee2018}
Lee,~W.-J.; Chang,~Y.-H. Growth without Postannealing of Monoclinic VO2 Thin
  Film by Atomic Layer Deposition Using VCl4 as Precursor. \emph{Coatings}
  \textbf{2018}, \emph{8}\relax
\mciteBstWouldAddEndPuncttrue
\mciteSetBstMidEndSepPunct{\mcitedefaultmidpunct}
{\mcitedefaultendpunct}{\mcitedefaultseppunct}\relax
\EndOfBibitem
\bibitem[Dieterle \latin{et~al.}(2002)Dieterle, Weinberg, and
  Mestl]{Dieterle2002}
Dieterle,~M.; Weinberg,~G.; Mestl,~G. Raman spectroscopy of molybdenum oxides
  Part I. Structural characterization of oxygen defects in MoO3-x by DR
  UV/VIS{,} Raman spectroscopy and X-ray diffraction. \emph{Phys. Chem. Chem.
  Phys.} \textbf{2002}, \emph{4}, 812--821\relax
\mciteBstWouldAddEndPuncttrue
\mciteSetBstMidEndSepPunct{\mcitedefaultmidpunct}
{\mcitedefaultendpunct}{\mcitedefaultseppunct}\relax
\EndOfBibitem
\bibitem[Hardcastle \latin{et~al.}(1988)Hardcastle, Wachs, Horsley, and
  Via]{Hardcastle1988}
Hardcastle,~F.~D.; Wachs,~I.~E.; Horsley,~J.~A.; Via,~G.~H. The structure of
  surface rhenium oxide on alumina from laser raman spectroscopy and x-ray
  absorption near-edge spectroscopy. \emph{Journal of Molecular Catalysis}
  \textbf{1988}, \emph{46}, 15--36\relax
\mciteBstWouldAddEndPuncttrue
\mciteSetBstMidEndSepPunct{\mcitedefaultmidpunct}
{\mcitedefaultendpunct}{\mcitedefaultseppunct}\relax
\EndOfBibitem
\bibitem[Weber and McGinnis(1960)Weber, and McGinnis]{Weber1960}
Weber,~A.; McGinnis,~E.~A. The Raman spectrum of gaseous oxygen. \emph{Journal
  of Molecular Spectroscopy} \textbf{1960}, \emph{4}, 195--200\relax
\mciteBstWouldAddEndPuncttrue
\mciteSetBstMidEndSepPunct{\mcitedefaultmidpunct}
{\mcitedefaultendpunct}{\mcitedefaultseppunct}\relax
\EndOfBibitem
\bibitem[Caylan and Cambaz~Buke(2021)Caylan, and Cambaz~Buke]{Caylan2021}
Caylan,~O.~R.; Cambaz~Buke,~G. Low-temperature synthesis and growth model of
  thin Mo2C crystals on indium. \emph{Sci. Rep.} \textbf{2021}, \emph{11},
  8247\relax
\mciteBstWouldAddEndPuncttrue
\mciteSetBstMidEndSepPunct{\mcitedefaultmidpunct}
{\mcitedefaultendpunct}{\mcitedefaultseppunct}\relax
\EndOfBibitem
\bibitem[Javey \latin{et~al.}(2003)Javey, Guo, Wang, Lundstrom, and
  Dai]{Javey2003}
Javey,~A.; Guo,~J.; Wang,~Q.; Lundstrom,~M.; Dai,~H. Ballistic carbon nanotube
  field-effect transistors. \emph{Nature} \textbf{2003}, \emph{424},
  654--657\relax
\mciteBstWouldAddEndPuncttrue
\mciteSetBstMidEndSepPunct{\mcitedefaultmidpunct}
{\mcitedefaultendpunct}{\mcitedefaultseppunct}\relax
\EndOfBibitem
\bibitem[Talirz \latin{et~al.}(2017)Talirz, Söde, Dumslaff, Wang,
  Sanchez-Valencia, Liu, Shinde, Pignedoli, Liang, Meunier, Plumb, Shi, Feng,
  Narita, Müllen, Fasel, and Ruffieux]{Talirz2017}
Talirz,~L.; Söde,~H.; Dumslaff,~T.; Wang,~S.; Sanchez-Valencia,~J.~R.;
  Liu,~J.; Shinde,~P.; Pignedoli,~C.~A.; Liang,~L.; Meunier,~V.; Plumb,~N.~C.;
  Shi,~M.; Feng,~X.; Narita,~A.; Müllen,~K.; Fasel,~R.; Ruffieux,~P.
  On-Surface Synthesis and Characterization of 9-Atom Wide Armchair Graphene
  Nanoribbons. \emph{ACS Nano} \textbf{2017}, \emph{11}, 1380--1388, PMID:
  28129507\relax
\mciteBstWouldAddEndPuncttrue
\mciteSetBstMidEndSepPunct{\mcitedefaultmidpunct}
{\mcitedefaultendpunct}{\mcitedefaultseppunct}\relax
\EndOfBibitem
\bibitem[Svensson and Campbell(2011)Svensson, and Campbell]{Svensson2011}
Svensson,~J.; Campbell,~E. E.~B. Schottky barriers in carbon nanotube-metal
  contacts. \emph{J. Appl. Phys.} \textbf{2011}, \emph{110}, 111101\relax
\mciteBstWouldAddEndPuncttrue
\mciteSetBstMidEndSepPunct{\mcitedefaultmidpunct}
{\mcitedefaultendpunct}{\mcitedefaultseppunct}\relax
\EndOfBibitem
\end{mcitethebibliography}

\end{document}







\setcounter{page}{1}
\renewcommand{\thepage}{S-\arabic{page}}

\maketitle 

\section{Temperature dependence of other devices}

\begin{figure}[H]
    \centering
    \begin{subfigure}[b]{0.4\linewidth}
        \phantomcaption{}
        \centering
        \includegraphics[scale=1]{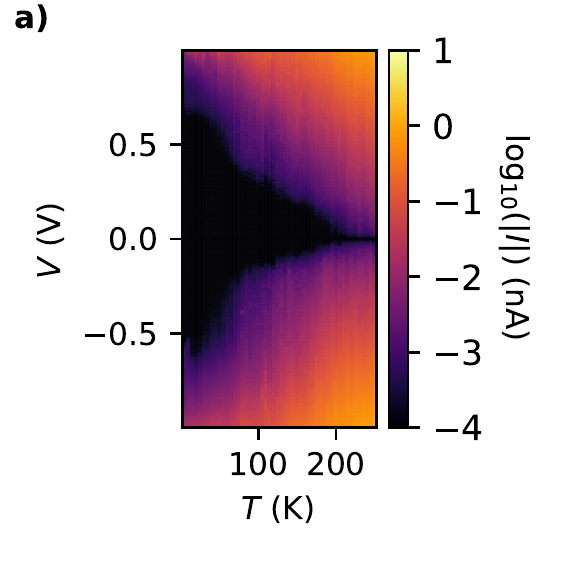}
        
        \label{fig:S1a}
    \end{subfigure}
    \begin{subfigure}[b]{0.4\linewidth}
        \phantomcaption{}
        \centering
        \includegraphics[scale=1]{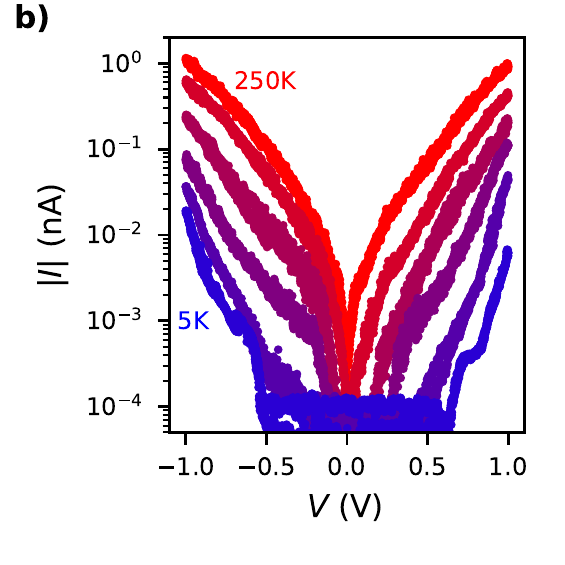}
        
        \label{fig:S1b}
    \end{subfigure}
    \begin{subfigure}[b]{0.4\linewidth}
        \phantomcaption{}
        \centering
        \includegraphics[scale=1]{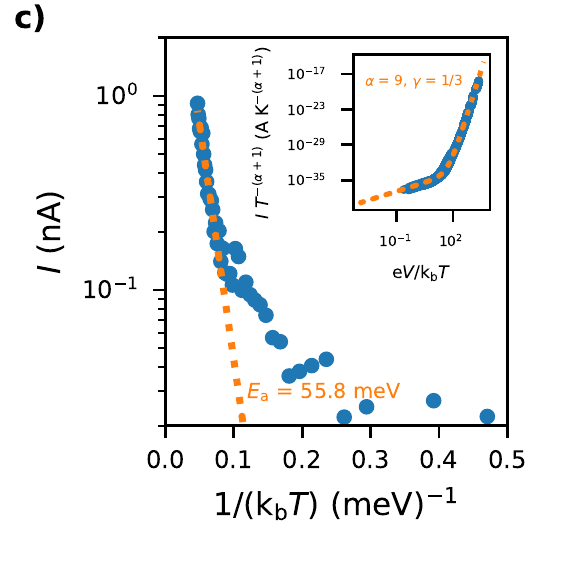}
        
        \label{fig:S1c}
    \end{subfigure}

    \caption{a) Map of current versus bias voltage and temperature of a selected wide MoRe nanogap 9-AGNR device.
    b) Corresponding temperature dependence of the current-voltage characteristic extracted from a).
    c) Corresponding temperature dependence of the current at $V$ = 1~V extracted from a). The inset shows a scaling analysis with a guide to the eye based on the nuclear tunneling model.
    }
    \label{fig:S1}
\end{figure}

In this section we show the $IV$ curves of devices which were measured versus temperature, but were not shown in the main text. Fig.~\ref{fig:S1a} shows the $IV$ characteristic of a wide MoRe nanogap 9-AGNR device versus temperature as a colormap. Individual $IV$ curves from this map are plotted in Fig.~\ref{fig:S1b}. In Fig.~\ref{fig:S1c}, the same $E_{\mathrm{a}}$ and nuclear tunneling scaling analysis that was done in the main text is performed on this dataset. The resulting activation energy is $E_{\mathrm{a}}$ = 60~meV, which is larger than the value found in the main text, but in the same order of magnitude. The scaling curve with the same parameters $\upalpha = 9$ and $\gamma = 3$ shows reasonable agreement. 

\begin{figure}[H]
    \centering
    \begin{subfigure}[b]{0.33\linewidth}
        \phantomcaption{}
        \centering
        \includegraphics[scale=1]{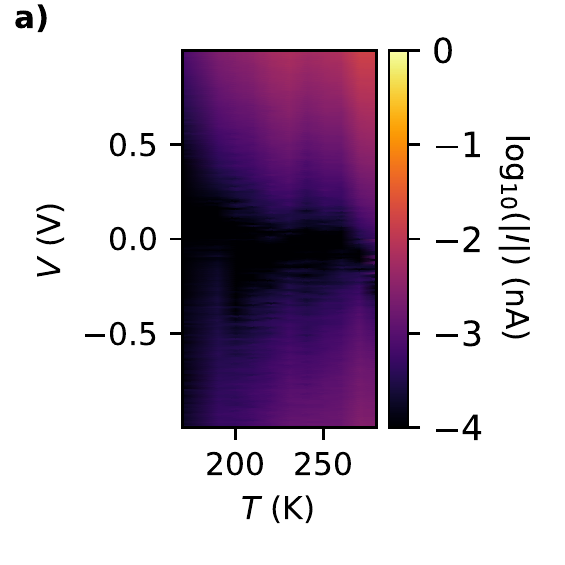}
        
        \label{fig:S2a}
    \end{subfigure}
    \begin{subfigure}[b]{0.33\linewidth}
        \phantomcaption{}
        \centering
        \includegraphics[scale=1]{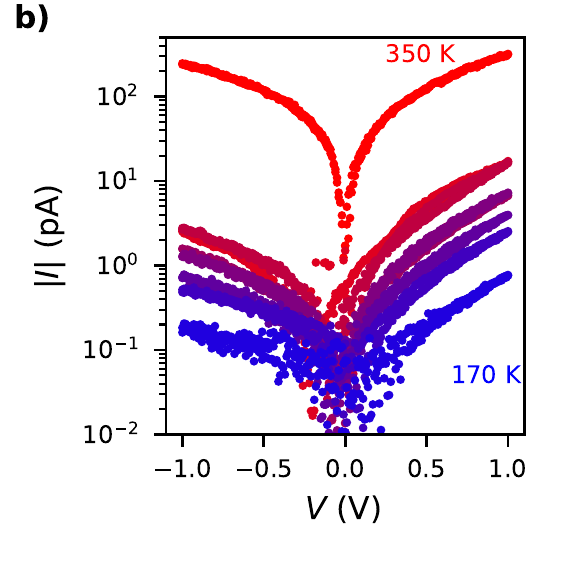}
        
        \label{fig:S2b}
    \end{subfigure}
    \begin{subfigure}[b]{0.33\linewidth}
        \phantomcaption{}
        \centering
        \includegraphics[scale=1]{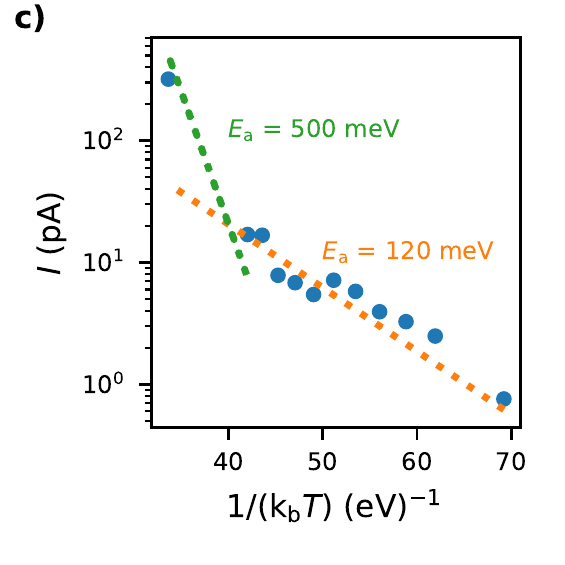}
        
        \label{fig:S2c}
    \end{subfigure}

    \caption{a) Map of current versus bias voltage and temperature of a selected 10-nm nanogap MoRe 9-AGNR device.
    b) Corresponding temperature dependence of the current-voltage characteristic extracted from b).
    c) Corresponding temperature dependence of the current at $V$ = 1~V extracted from b). The inset shows a scaling analysis with a guide to the eye based on the nuclear tunneling model.
    }
    \label{fig:S2}
\end{figure}

$IV$ curves for another 10-nm nanogap MoRe 9-AGNR device (represented by the blue curve in Fig.~3b in the main text) were taken in the temperature range of 280~K to 170~K at a bottom-gate voltage of -2~V. The data is shown as a colormap in Fig.~\ref{fig:S2a}. The current-voltage curves from this colormap are plotted in Fig.~\ref{fig:S2b}, alongside an $IV$ curve that was taken at $T=$~350~K. In Fig.~\ref{fig:S2c} the current at 1 V bias voltage and -2~V gate voltage is plotted versus inverse temperature $\frac{1}{\mathrm{k_{b}}T}$. The activation energy fits from Fig.~3f are overlaid onto this plot, showing reasonable agreement with the data. 

\begin{figure}[H]
    \centering

    \includegraphics[scale=1]{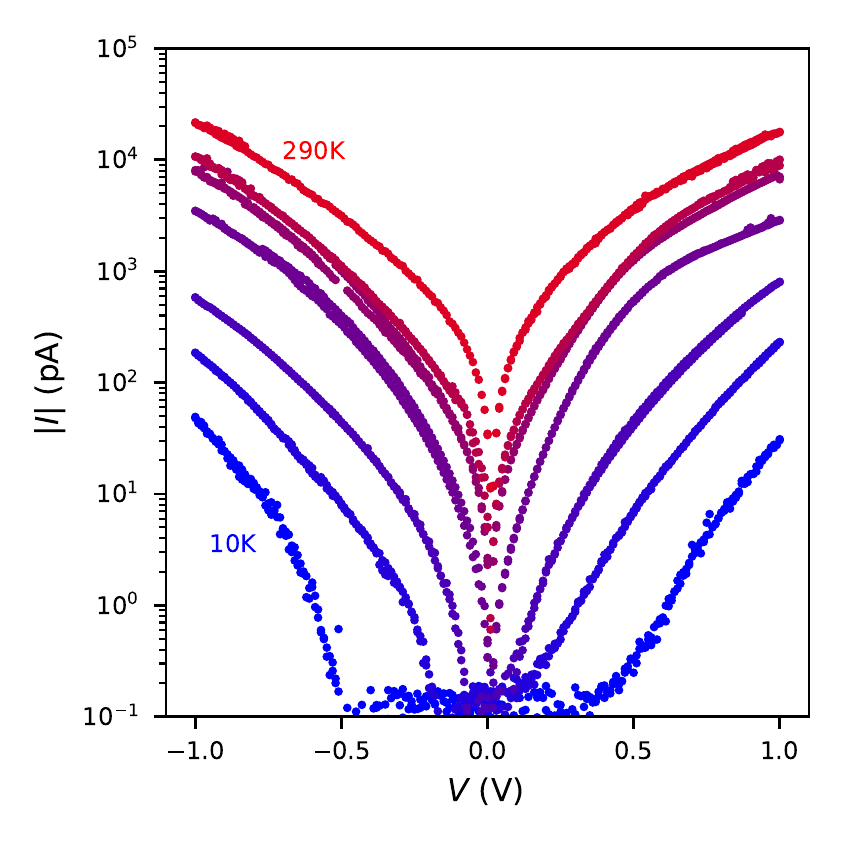}
    
    \caption{Current versus bias voltage and temperature of a selected 10-nm nanogap Pd 9-AGNR device at temperatures 10, 70, 100, 150, 200, 220 and 290~K.
    }
    \label{fig:S3}
\end{figure}

The $IV$ curve of the other 10-nm nanogap Pd 9-AGNR device was taken at temperatures T = 12, 72, 100, 150, 200, 220, 290~K. The resulting curves are shown in Fig.~\ref{fig:S3}.

\section{$IV$ to 4~V}

\begin{figure}[H]
    \centering
    \includegraphics{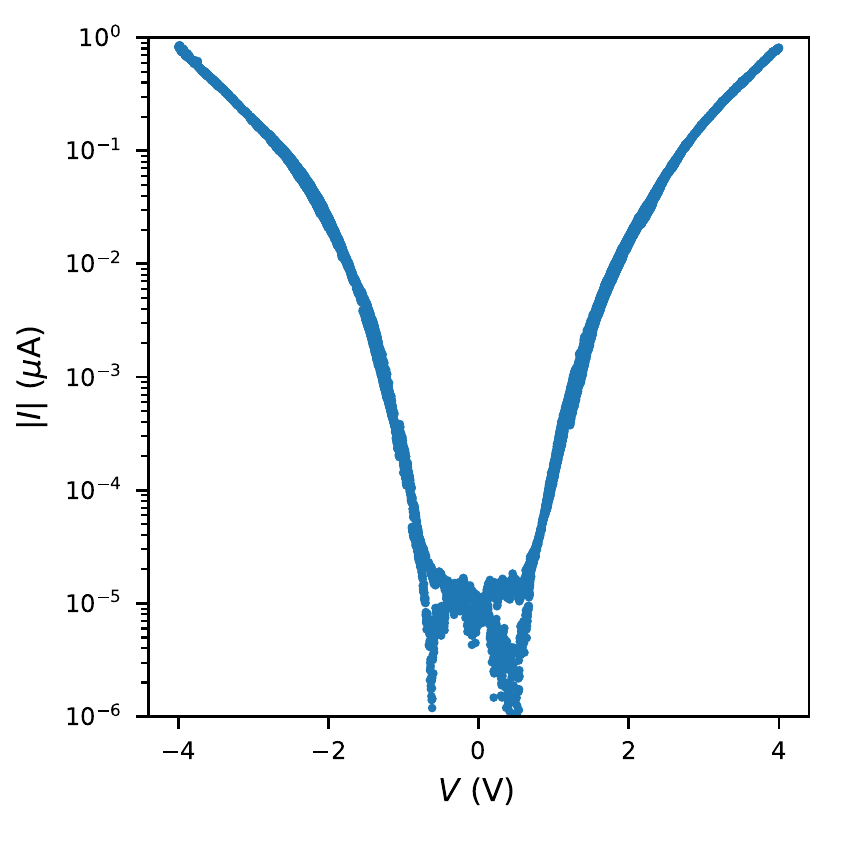}

    \caption{$IV$ curve up to a bias voltage of 4~V of a wide MoRe nanogap 9-AGNR device at a temperature of 100~mK}
    \label{fig:S4}
\end{figure}

In Fig.~\ref{fig:S4}, we show the $IV$ curve of a wide nanogap MoRe 9-AGNR device up taken up to 4~V bias voltage, taken at a base temperature of 100~mK. The $IV$ characteristic remains highly nonlinear at higher bias voltages, reaching approximately 1~$\upmu$A at a bias voltage of 4~V.

\section{\textcolor{black}{Asymmetry of the} $IV$ characteristic of 10-nm nanogap MoRe 9-AGNR devices}

\begin{figure*}[!ht]
    \centering
    \includegraphics{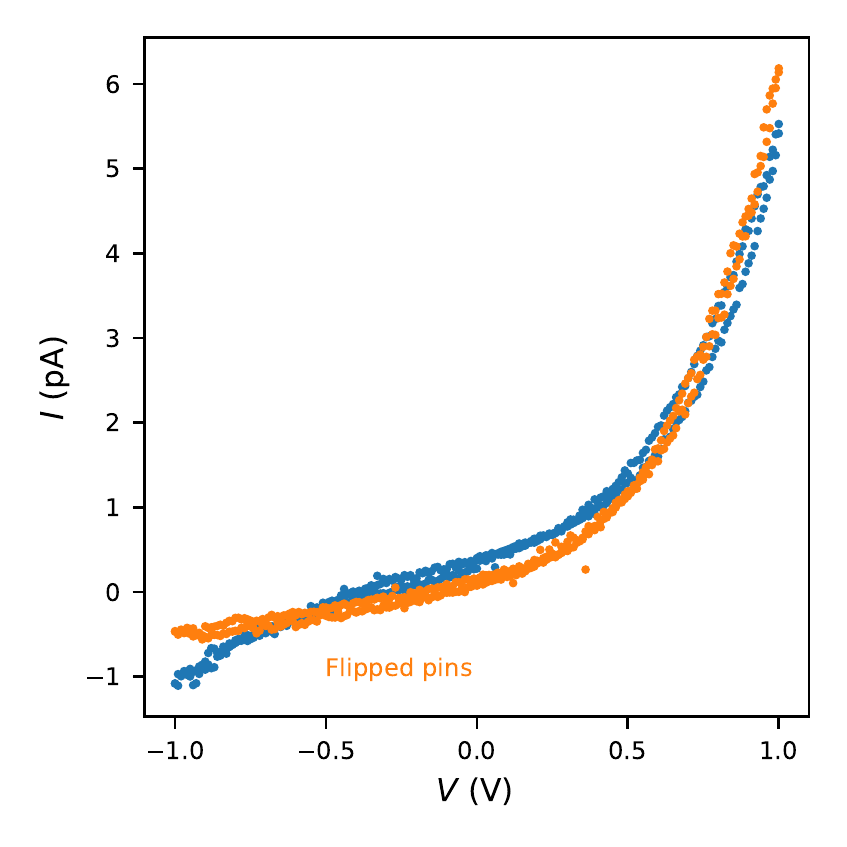}

    \caption{Current-voltage curve of the 10-nm nanogap MoRe 9-AGNR device studied in the main text. The blue curve is the \textcolor{black}{blue} $IV$ curve shown \textcolor{black}{in Fig.~3b)} in the main text. The $IV$ curve upon interchanging source and drain connections ('Flipped pins') is plotted in orange.}
    \label{fig:S10}
\end{figure*}

The asymmetry of the $IV$ curves of the 10-nm nanogap MoRe nanogap 9-AGNR device discussed in the main text in Fig.~3c-f was further studied. The source and drain connections are interchanged (which we denote by `Flipped pins'), which reverses the bias direction while preserving the average of the source and drain voltage with respect to the gate voltage. The $IV$ curve with flipped pins was taken at a gate voltage of $-2$~V, shown in Fig.~\ref{fig:S10} together with the $IV$ curve without the interchanged connections. Both $IV$ curves are asymmetric, with a larger current at positive bias voltage. \textcolor{black}{This suggests that not just $V_{\mathrm{source}}-V_{\mathrm{drain}}$, but also $(V_{\mathrm{source}}+V_{\mathrm{drain}})/2$ affects the $IV$ characteristic of the device.} A possible explanation for this could be an effective field effect from the voltage bias on the source and drain electrodes. \textcolor{black}{By applying a positive source or drain voltage, the effective back gate voltage $V_{gate}-\alpha (V_{\mathrm{source}}+V_{\mathrm{drain}})/2$, where $\alpha$ is a constant proportional to the ratio of source and drain capacitance to gate capacitance, becomes more negative}\textcolor{black}{. This} effectively \textcolor{black}{p-dopes (n-dopes)} the GNR \textcolor{black}{channel for positive (negative) bias voltages}. Since this effect does not depend on the bias direction, it is unaffected by interchanging the source and drain pins. \textcolor{black}{A small asymmetry remains upon flipping the pins, which could be due to a small asymmetry between source and drain capacitance or possibly due to asymmetry in the electronic coupling.}

\section{Time dependence of hysteresis effect}

To investigate the time-dependence of the hysteresis effect, two additional types of measurement were done on the 10-nm nanogap MoRe 9-AGNR device characterized in the main text.

\begin{figure}[H]
    \centering
    \includegraphics{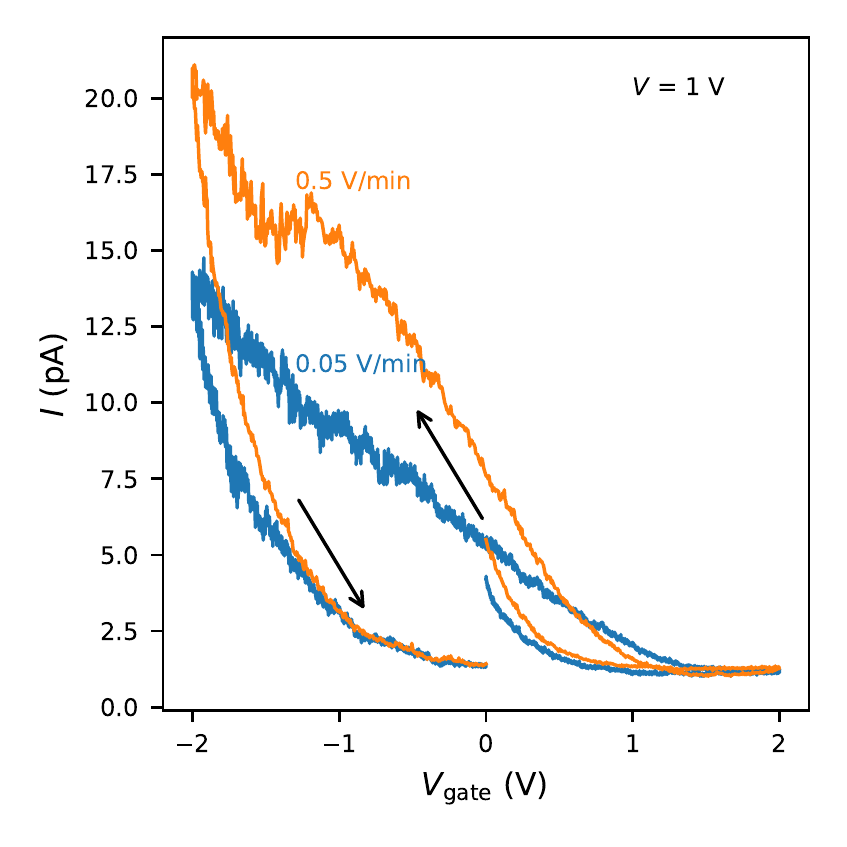}

    \caption{Current versus gate voltage for the 10-nm nanogap MoRe 9-AGNR device in the main text. The curve was taken at two different ramping speeds, $0.5$~V/min and $0.05$~V/min. The arrows indicate the sweep direction of the gate voltage.}
    \label{fig:S8}
\end{figure}

In the first measurement, the current was measured versus gate voltage at a fixed bias voltage of 1~V. The measurement was performed at two different gate voltage ramp speeds, $0.5$~V/min and $0.05$~V/min. The resulting curves are shown in Fig.~\ref{fig:S8}.
Both curves display hysteresis, with a larger current loop for the faster sweep. The dependence on the sweep rate indicates that the hysteresis loop is time dependent. 

\begin{figure}[H]
    \centering
    \includegraphics{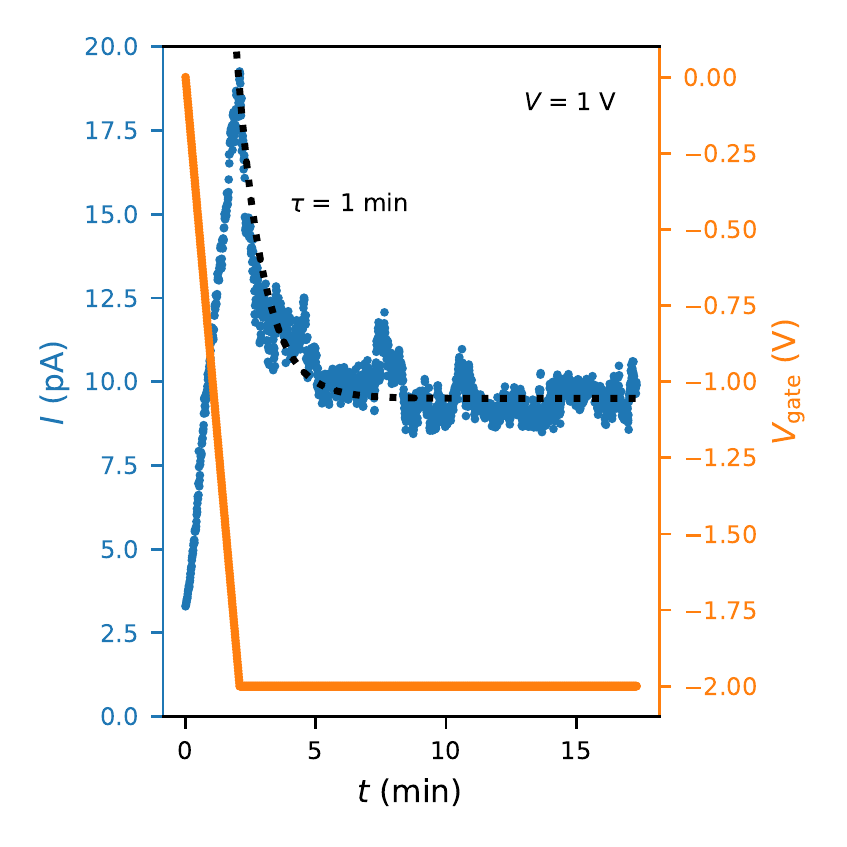}

    \caption{Measurement of the current versus time during and after ramping the gate-voltage to $-2$~V at a rate of $1$~V/min. The bias voltage was fixed at $1$~V. The current(gate voltage) is plotted in blue(orange). As a guide to the eye, an offset exponential decay with time constant $\tau=1~\mathrm{min}$ is plotted(dotted black).}
    \label{fig:S9}
\end{figure}

A second type of measurement was performed to further characterize this time-dependence, the current was measured versus time during - and after - the ramping of the gate voltage to $-2$~V at a rate of $1$~V/min. The resulting data is shown in Fig.~\ref{fig:S9}. The blue curve in this figure shows the current versus time, while the orange curve shows the gate voltage versus time. The dotted black curve is a guide to the eye, which shows an exponential dependence $I=I_0+I_1e^{-\frac{t}{\tau}}$ with $\tau = 1~\mathrm{min}$. As the gate voltage is ramped, the current increases without a delay. From the end of the gate voltage ramp at $2$ minutes onward, the current is well-described by an exponential decay with a timescale of roughly $1$ minute. The time-effect thus counteracts the field effect from the back-gate. This suggests a charge-transfer mechanism~\cite{Wang2010}, which may originate from electrochemical reactions with adsorbents, such as water~\cite{Kim2003,Xu2012}\textcolor{black}{, in which silanol groups at the SiO$_2$ substrate termination are known to play a role}. The hysteresis is opposite from the previously reported hysteresis observed in GNR devices attributed to charge traps in SiO$_2$~\cite{Tries2020}.\textcolor{black}{Hysteresis due to adsorbents has previously been reduced by passivation of devices with HMDS\cite{Bennett2013} to make the substrate hydrophobic. For a more detailed discussion on hysteresis effects in graphene nanostructures, we refer to the review article by Lu et al.\cite{Lu2022}.}

\section{Nuclear tunneling analysis of the 10-nm nanogap Pd 9-AGNR device}

In Fig.~S11, we show a nuclear tunneling scaling analysis of the data in Fig.~4a. By eye, the rescaled $IV$ curves all fall onto a single curve for $\alpha=5$. A fit was done for $\gamma$, which resulted in $\gamma=0.145$, which corresponds to roughly $7$ hopping sites. We note that the curve does not fit well for $\frac{eV}{k_{\mathrm{b}}T}<10$. Furthermore, the number of hopping sites is large given the $6$~nm contact spacing. This suggests that the nuclear tunneling model is not a satisfactory description of the electronic transport characteristics in this device.  

\begin{figure*}[!ht]
    \centering
    \includegraphics{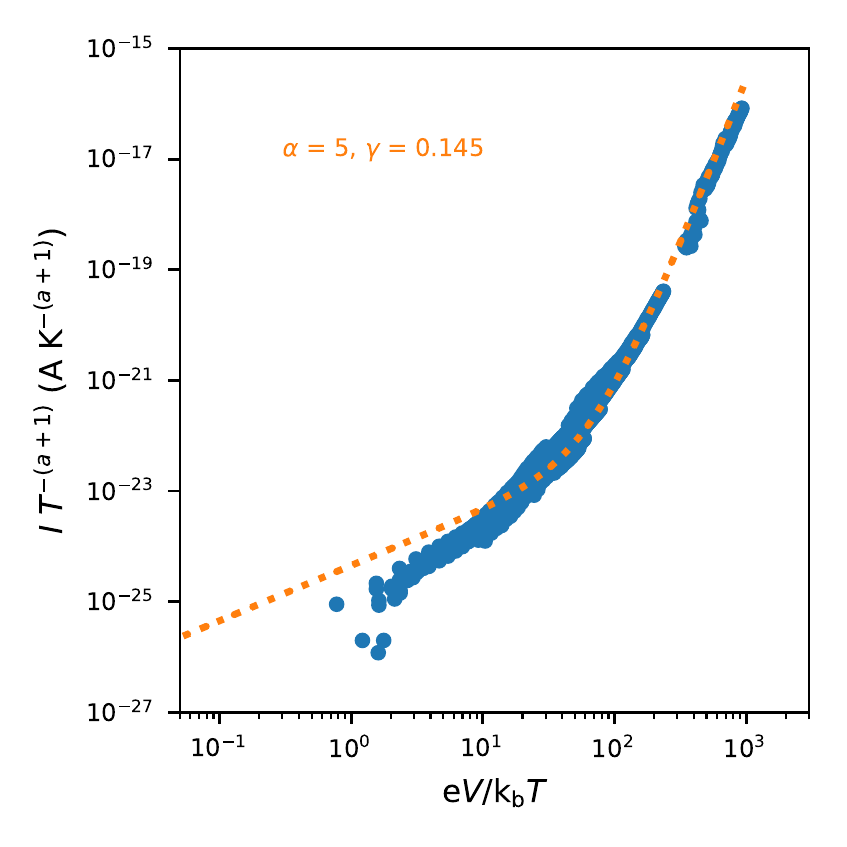}

    \caption{Nuclear tunneling scaling analysis for the selected 10-nm nanogap Pd 9-AGNR device.}
    \label{fig:S11}
\end{figure*}

\section{Raman Spectroscopy of GNR devices}

\begin{figure}[H]
    \centering
    \includegraphics[scale=0.90]{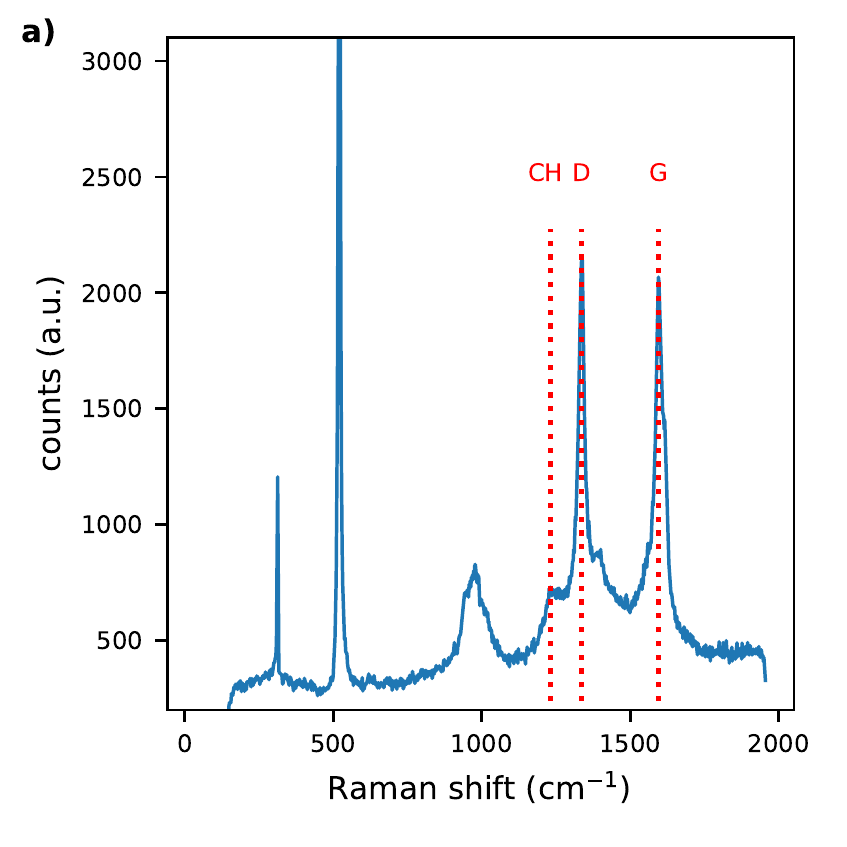}
    \includegraphics[scale=0.90]{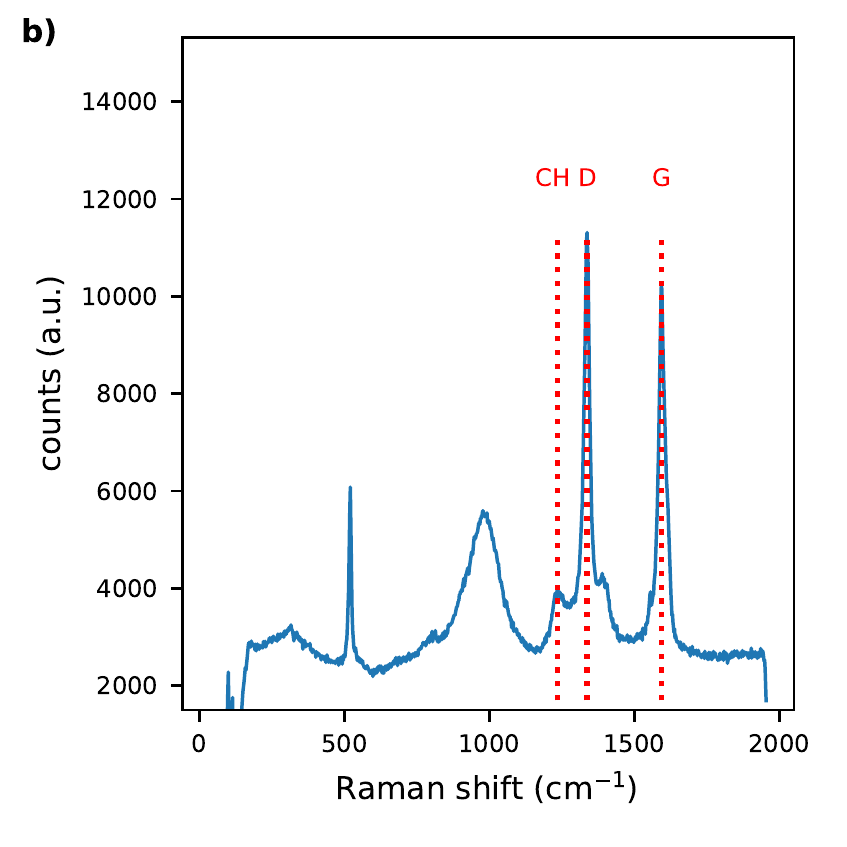}

    \caption{\textcolor{black}{Raman spectrum taken with a 514 nm laser (a) on the SiO$_2$ near the wide MoRe nanogap 9-AGNR devices (b) on the 15 nm HfO$_2$ covered TiPt bottom gate near the needle-like MoRe nanogap 9-AGNR devices. The G, D and CH modes are indicated by the red dotted lines.}}
    \label{fig:S12}
\end{figure}

\textcolor{black}{To confirm the presence and structural integrity of the GNRs, Raman spectroscopy ($\lambda$ = 514 nm) was performed on the GNRs for the wide MoRe nanogap electrodes and the needle-like MoRe nanogap electrodes. In the case of the wide MoRe nanogap electrodes, characterization was performed of GNRs on the 285 nm thick SiO$_2$. For the needle-like MoRe electrodes, the characterization was performed on the 15 nm HfO$_2$ covered TiPt bottom gate, which acts as a Raman optimized substrate\cite{Overbeck2019}. The resulting Raman spectra are shown in Fig.~\ref{fig:S12}. The energies of the characteristic G, D and CH peaks for the polymer-free transferred method are 1595 cm$^{-1}$, 1337 cm$^{-1}$ and 1230 cm$^{-1}$ respectively. For the PMMA transfer, the G, D and CH peaks are found at 1593 cm$^{-1}$, 1337 cm$^{-1}$ and 1234 cm$^{-1}$. This is similar to the spectra that have been measured before for 9-AGNRs \cite{BorinBarin2019, Overbeck2019}. The presence of the G peak around 1595 cm$^{-1}$ in both spectra suggests a limited degree of doping in the transferred GNRs and no significant difference between the two substrate transfer methods. Beside the G, D and CH peaks, a sharp peak at 313 cm$^{-1}$ for the wide nanogap GNR devices and a broad peak at 311 cm$^{-1}$ for the needle-like GNR devices can be indicative of the radial breathing-like mode (RBLM), which is related to the width of the 9-AGNRs, expected at 311 cm$^{-1}$. Although the observed wavenumber matches the expected RBLM wavenumber, it should be noted that Si also has a Raman active vibration TA mode at 301 cm$^{-1}$\cite{Lee2018}. Thus we can not with certainty identify the broad peak in the needle-like GNR devices as the RBLM peak. If we suppose that it is the RBLM peak, a possible explanation for the broadening of the RBLM peak could be increased damping of large wavelength vibrations in the GNRs transferred by the PMMA-assisted method. The peaks at 520 cm$^{-1}$ and 950 cm$^{-1}$ are related to vibrational modes of the silicon substrate.}

\begin{figure}[H]
    \centering
    \includegraphics[scale=0.90]{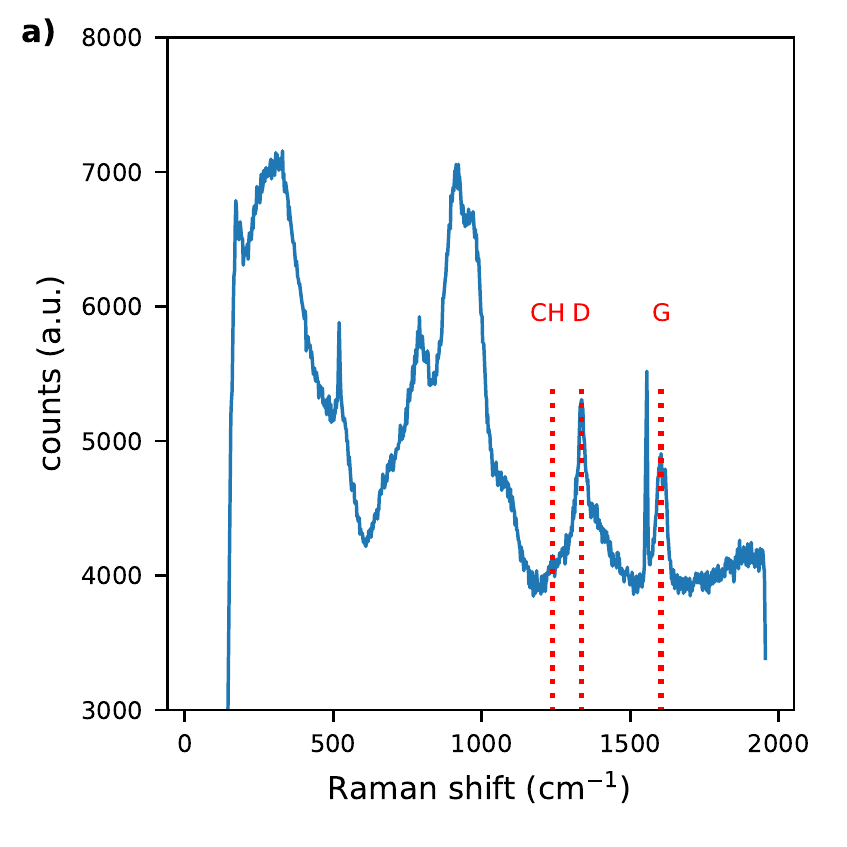}
    \includegraphics[scale=0.90]{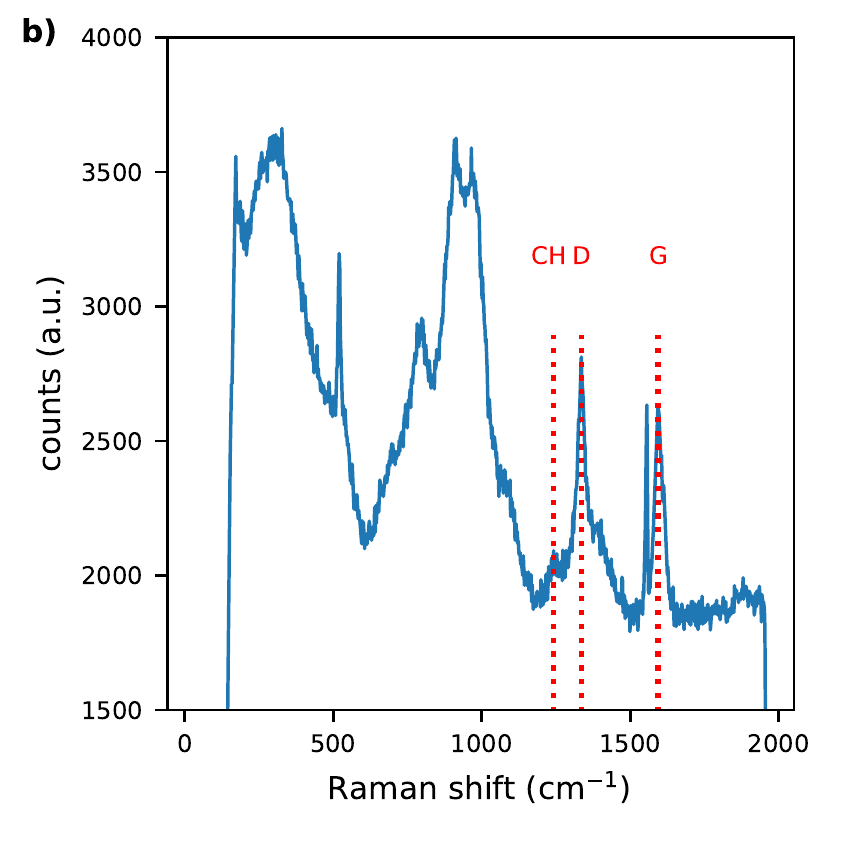}
    \includegraphics[scale=0.90]{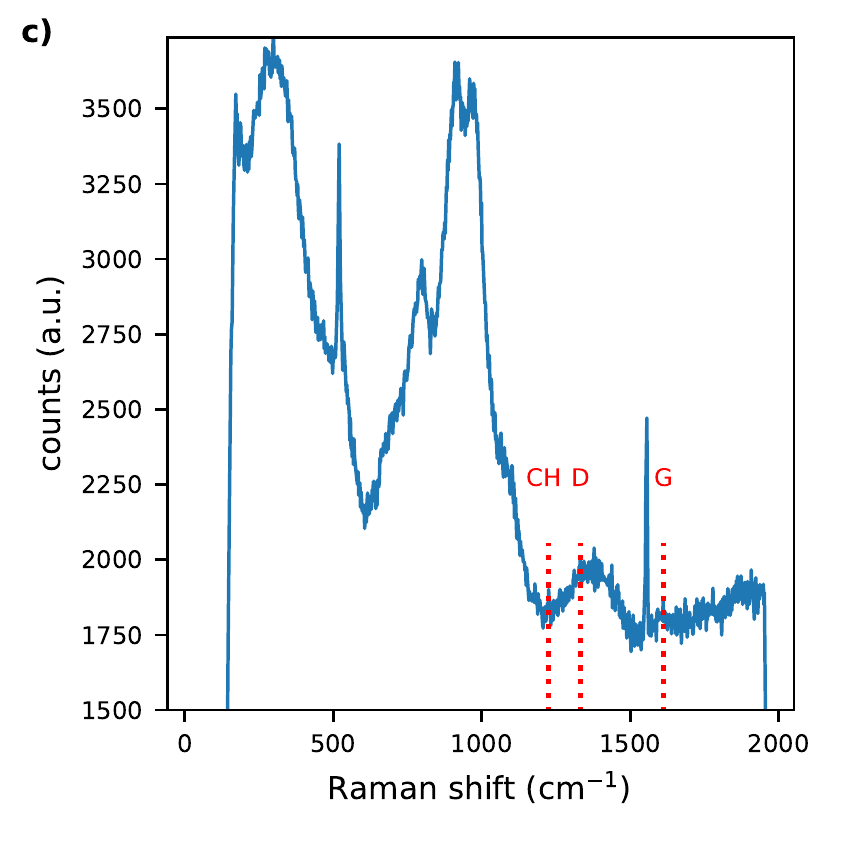}

    \caption{\textcolor{black}{Raman spectrum taken with a 514 nm laser (a) on the contact pad of a wide MoRe nanogap 9-AGNR device (b) on the contact pad of a needle-like MoRe nanogap 9-AGNR device. c) on the contact pad of a needle-like MoRe nanogap device that was not covered by GNRs. The identified G, D and CH modes are indicated by the red dotted lines.}}
    \label{fig:S14}
\end{figure}

\textcolor{black}{To investigate the properties of GNRs on MoRe as well, Raman spectra were taken on the MoRe part of the devices. In Fig.~\ref{fig:S14}, we show Raman spectra on the contact pads of the wide MoRe nanogap devices and the needle-like MoRe nanogap devices. The C-H, D and G peaks show the presence of the GNRs on the MoRe. The Raman peaks between 800 cm$^-{1}$ and $1000$ cm$^{-1}$ are characteristic for the vibrational modes of Mo=O bonds\cite{Dieterle2002} and Re=O bonds\cite{Hardcastle1988}. In addition, we observe an additional peak below the G peak of the GNRs, at a Raman shift of 1556 cm$^{-1}$.}

\textcolor{black}{To investigate the new peak on MoRe close to the G-peak, a Raman spectrum was taken on an MoRe contact pad without GNRs. This spectrum is shown in Fig.~\ref{fig:S14} c). In this spectrum, we still observe the presence of the vibrational peak at 1556 cm$^{-1}$, which suggests that it should not be attributed to interaction of MoRe and GNRs, but rather to a vibrational mode related to MoRe itself. The peak is rather sharp and appears to correspond to molecular $O_2$\cite{Weber1960}, suggesting adsorption of $O_2$ gas onto the MoRe film or release of O$_2$ gas by the MoRe film during the Raman spectroscopy. By comparing the spectrum on bare MoRe to the spectrum with GNRs, We do not observe any obvious peaks that might be related to the formation of Mo-C bonds, which are expected to be found at 231 and 656 cm$^{-1}$\cite{Caylan2021}. We do however see that the broad peak around 300 cm$^{-1}$ changes shape between the spectra with and without GNRs, again suggesting the presence of the RBLM mode.}

\section{Gate voltage dependence at $T$ = 350~K}

\begin{figure}[H]
    \centering
    \includegraphics{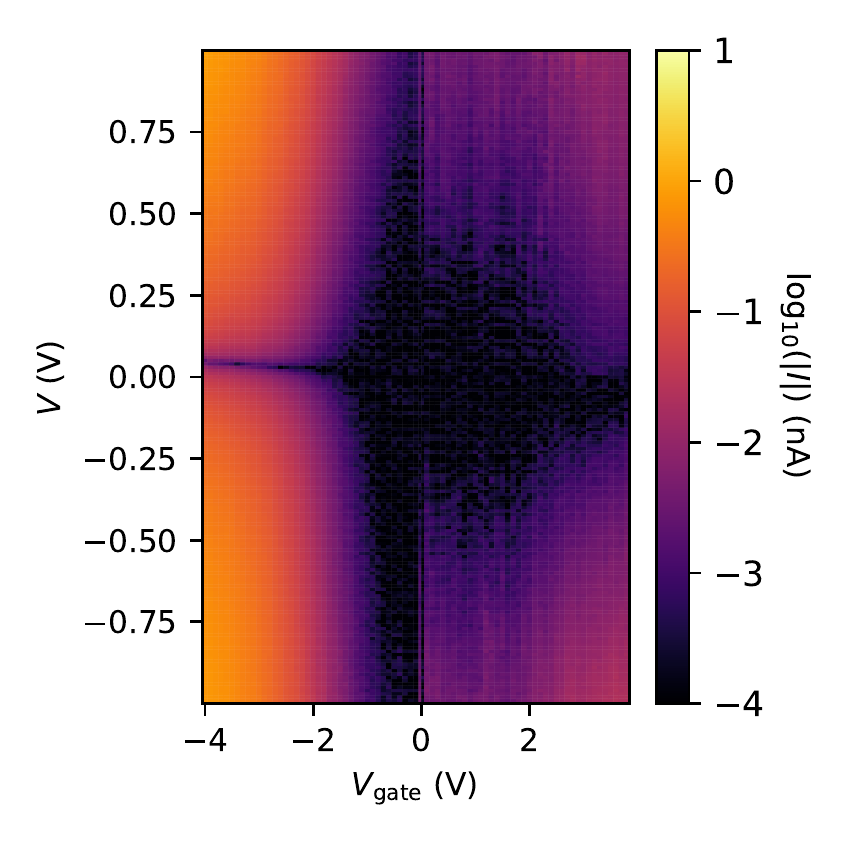}

    \caption{Logarithmic color map of the current versus bias voltage and gate voltage for a 10-nm nanogap MoRe 9-AGNR device at a temperature of 350~K}
    \label{fig:S5}
\end{figure}

\begin{figure}[H]
    \centering
    \includegraphics{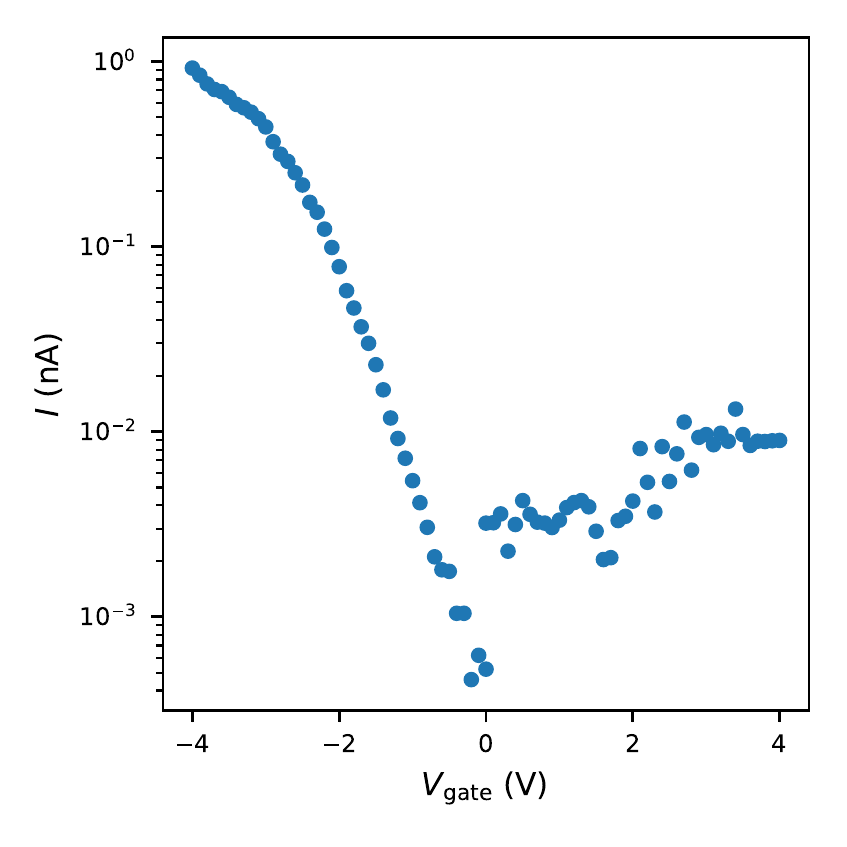}

    \caption{Current versus gate voltage curve at a fixed bias voltage of 1~V; extracted from Fig.~\ref{fig:S5}.}
    \label{fig:S6}
\end{figure}

For the 10-nm nanogap MoRe devices, we also studied the gate dependence at a temperature of $350$~K. This temperature was chosen because the measured conductance of the devices was larger at this temperature. We measured the current through the devices versus gate and bias voltage. In an attempt to reduce the effect of hysteresis observed in the main text, the data was first taken from a gate voltage of $0$~V to a gate voltage of $-4$~V and subsequently ramped back to $0$~V. After waiting for 10 minutes at $0$~V, another map was taken from 0~V to 4~V. The data is plotted in Figure~\ref{fig:S5} as a single color map. The color map does show a discontinuity at zero gate voltage. The device has a much larger current for negative gate voltages than for positive gate voltages. To further illustrate this, in Fig.~\ref{fig:S6} we plot a trace at a bias voltage of $1$~V, which shows that going from $0$~V gate voltage  to $-4$~V, the current increases by $3$ orders of magnitude, while an increase of at most $1$ order of magnitude is seen going from $V_{\mathrm{gate}}=0$~V to $V_{\mathrm{gate}}=4$~V. This is in line with the idea that the contact is p-type in nature. At gate voltages lower than $-3$~V, the increase in $\mathrm{log}(I)$ starts to flatten off.
The shape of the $IV$ characteristic of the devices also changes as a function of the applied gate voltage. 

\begin{figure}[H]
    \centering
    \includegraphics{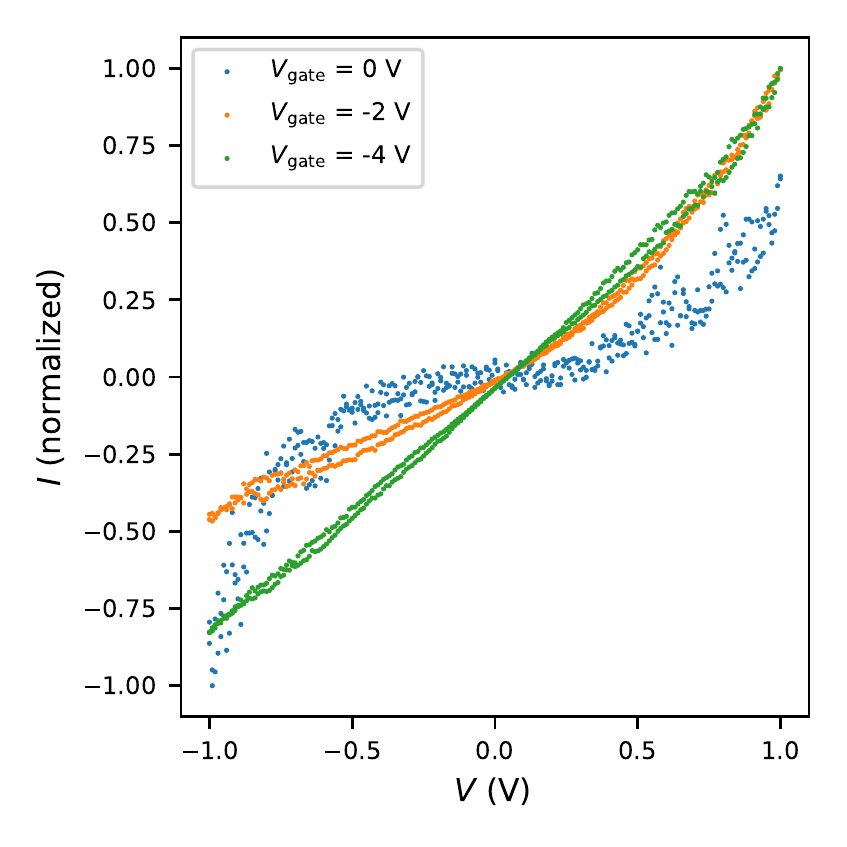}

    \caption{Normalized $IV$ curves of a 10-nm nanogap MoRe 9-AGNR at a temperature of 350~K and gate voltage of 0, -2, -4~V.}
    \label{fig:S7}
\end{figure}

In Fig.~\ref{fig:S7} we plot the $IV$ characteristic at gate voltages of $0$~V, $-2$~V and $-4$~V. The curves are normalized to $\mathrm{max}(|I|)$ to illuminate the change in shape. The $IV$ characteristic becomes more linear as the gate voltage is swept to negative values. At $0$~V gate voltage the $IV$ characteristic is highly non-linear, while at $-4$~V a nearly linear curve is obtained. In the transition region, the $IV$ characteristic is more bias voltage asymmetric, with a larger current for positive bias voltages. The increase in linearity could tentatively be explained by a change in the contact to a near-ohmic-regime. A current of only $1$~nA at $1$~V is however orders of magnitude below the conductance of the order of $G_0\simeq 77$~$\upmu$S observed for devices with ohmic contacts to carbon nanotubes~\cite{Javey2003}. In combination with the large activation energies for 10-nm nanogap MoRe devices, this highlights that the nature of the contact to the GNRs can not be determined solely from the linearity of the current-voltage characteristic.

\section{Band-alignment of 9-AGNRs with MoRe}
\begin{figure}[H]
    \centering
    \includegraphics{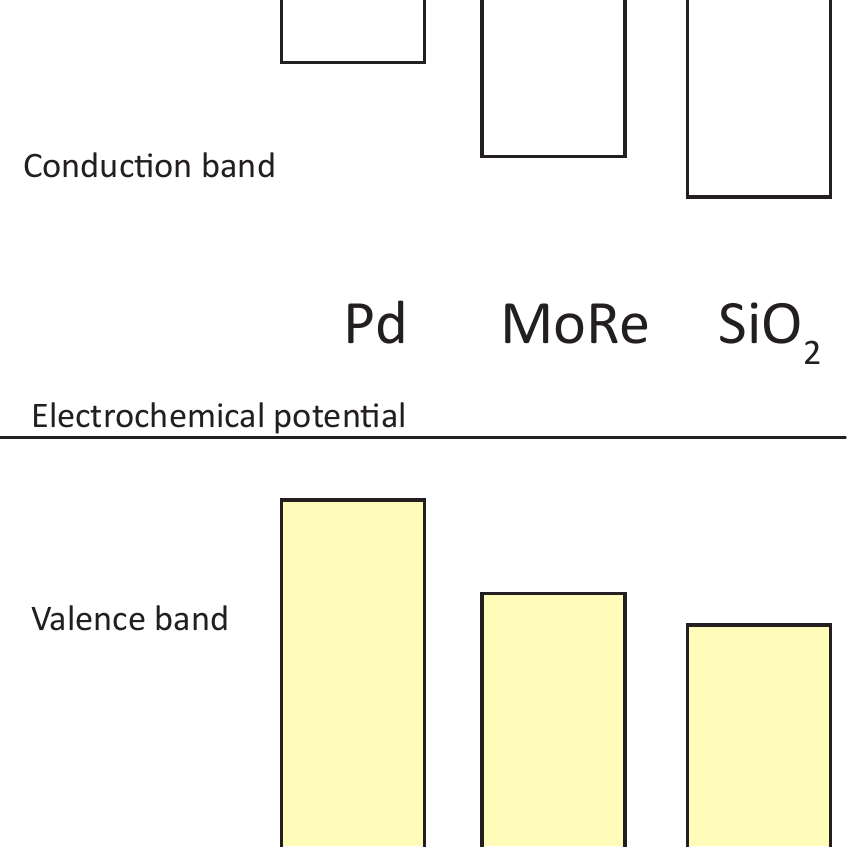}

    \caption{\textcolor{black}{Graphical representation of the band-alignment between 9-AGNR valence and conduction bands with respect to the chemical potential on MoRe, Pd and SiO$_2$ estimated by the Schottky-Mott rule.}}
    \label{fig:S15}
\end{figure}

\textcolor{black}{
In order to better understand the band alignment of 9-AGNRs with MoRe and Pd, we here illustrate the alignment of the conduction and valence band of 9-AGNRs with the electrochemical potential on MoRe, Pd and SiO$_2$, assuming an electronic band gap of $1.4$~eV\cite{Talirz2017} for metal adsorbed GNRs and a chemical potential of $4.6$~eV for 9-AGNRs in vacuum. The resulting band alignment is shown in Fig.~\ref{fig:S15}.  Following the main text, for MoRe, we take a chemical potential of $4.8$~eV and for Pd, we take a chemical potential of $5.1$ eV. We apply the Schottky-Mott rule to estimate the band alignment\cite{Svensson2011}, which is known to be valid for 9-AGNRs on Au\cite{Talirz2017} For the SiO$_2$, we assume no doping here. It should be noted that the actual band alignments could differ due to doping effects at the interfaces. Furthermore, the band gap of GNRs on an insulating substrate such as SiO$_2$ could differ from the band gap on a metal due to the lack of image charge effects on SiO$_2$\cite{Talirz2017}.}


\section{Optical microscopy images of the devices}

\begin{figure}[H]
    \centering
    \includegraphics[scale=1.0]{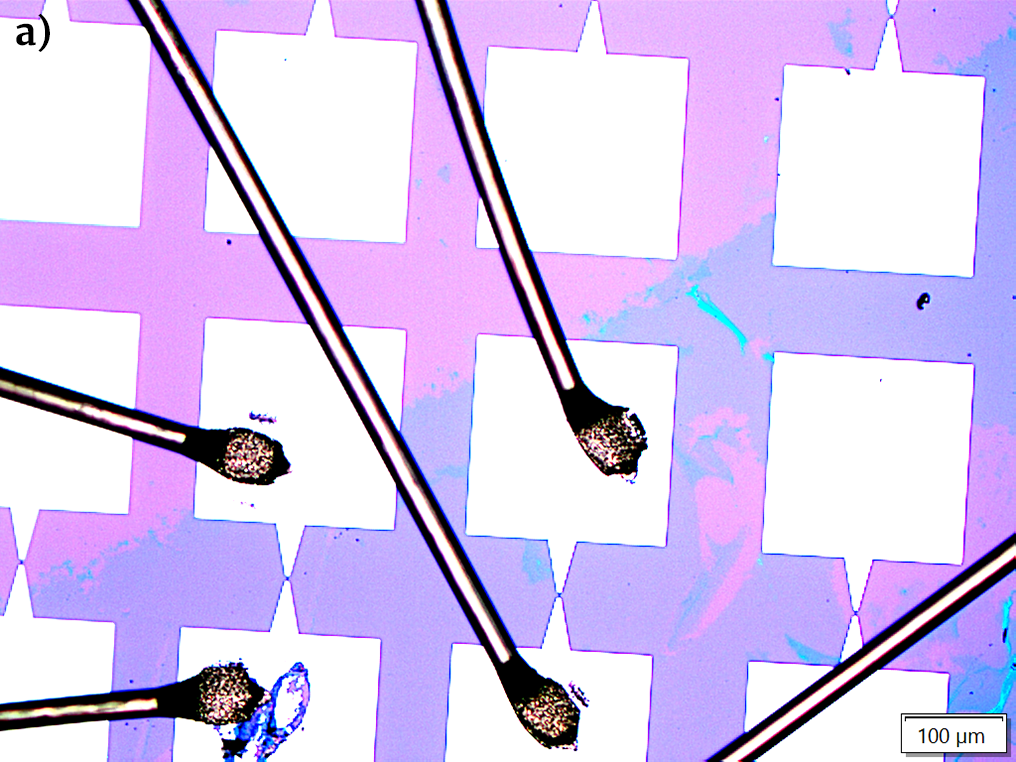}

    \includegraphics[scale=1.0]{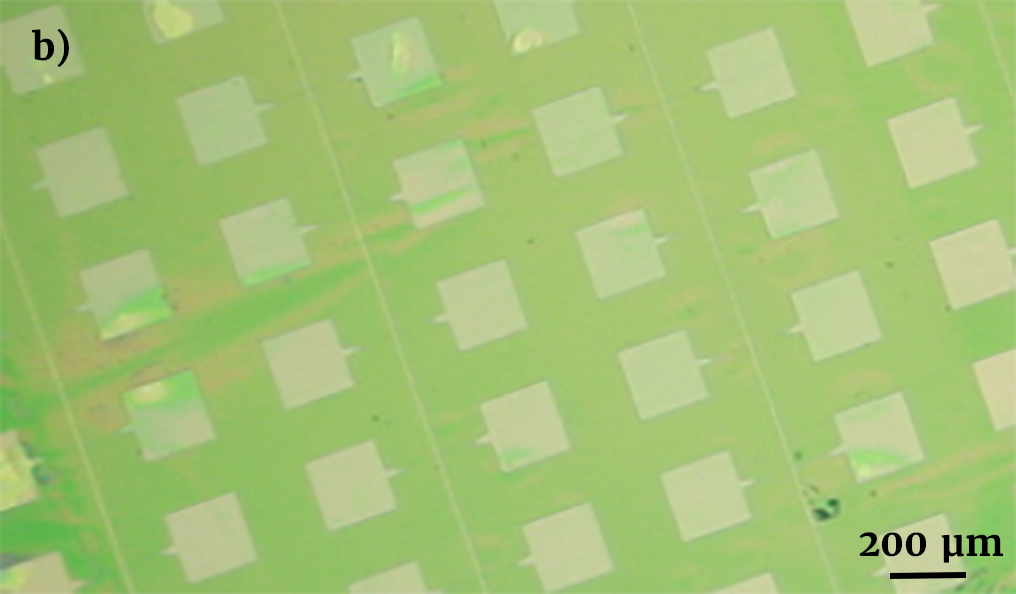}

    \caption{\textcolor{black}{Optical microscopy image of the 9-AGNR devices. a) Wide nanogap 9-AGNR MoRe devices with wirebonds. The GNR film is visible as a color change in the purposely oversaturated image. b) PMMA covered needle-like MoRe nanogap devices post PMMA-membrane assisted 9-AGNR transfer.} }
    \label{fig:S16}
\end{figure}

\textcolor{black}{For completeness, we here add optical microscopy images of the wide MoRe nanogap devices and the needle-like MoRe nanogap devices, which we show in Fig.~\ref{fig:S16}. In the wide MoRe nanogap devices, the 9-AGNR film can be seen as a discoloration on 285 nm thick SiO$_2$ on Si by optical microscopy images taken with an increased exposure time. The PMMA covered needle-like nanogap 9-AGNR devices on the other hand show no clearly visible GNR film.}

\bibliography{SI}